\documentclass[preprint,12pt]{elsarticle}

\usepackage{graphicx}
\usepackage{amsmath,bm}
\usepackage[table,xcdraw]{xcolor}
\usepackage[toc,page]{appendix}
\usepackage{natbib}
\biboptions{sort&compress}

\usepackage{placeins}
\makeatletter
\AtBeginDocument{%
  \expandafter\renewcommand\expandafter\subsection\expandafter
    {\expandafter\@fb@secFB\subsection}%
  \newcommand\@fb@secFB{\FloatBarrier
    \gdef\@fb@afterHHook{\@fb@topbarrier \gdef\@fb@afterHHook{}}}%
  \g@addto@macro\@afterheading{\@fb@afterHHook}%
  \gdef\@fb@afterHHook{}%
}
\makeatother

\usepackage{graphicx,epsfig,amssymb,amsmath,amsthm,amscd,amsfonts,bm}

\usepackage{amssymb}
\usepackage{amsmath}

\journal{arXiv}

\begin{document}

\begin{frontmatter}



\title{Hamiltonian-based neural networks for systems under nonholonomic constraints}


\author[inst1]{Ignacio Puiggros T.}

\author[inst1]{A. Srikantha Phani}

\affiliation[inst1]{organization={Department of Mechanical Engineering, The University of British Columbia},
            addressline={2054-6250 Applied Science Lane}, 
            city={Vancouver},
            postcode={V6T 1Z4}, 
            state={British Columbia},
            country={Canada}}

\begin{abstract}
There has been increasing interest in methodologies that incorporate physics priors into neural network architectures to enhance their modeling capabilities. A family of these methodologies that has gained traction are Hamiltonian neural networks (HNN) and their variations. These architectures explicitly encode Hamiltonian mechanics both in their structure and loss function. Although Hamiltonian systems under nonholonomic constraints are in general not Hamiltonian, it is possible to formulate them in pseudo-Hamiltonian form, equipped with a Lie bracket which is almost Poisson. This opens the possibility of using some principles of HNNs in systems under nonholonomic constraints.
The goal of the present work is to develop a modified Hamiltonian neural network architecture capable of modeling Hamiltonian systems under holonomic and nonholonomic constraints. A three-network parallel architecture is proposed to simultaneously learn the Hamiltonian of the system, the constraints, and their associated multipliers. A rolling disk and a ball on a spinning table are considered as canonical examples to assess the performance of the proposed Hamiltonian architecture. The experiments are then repeated with a noisy training set to study modeling performance under more realistic conditions.

\end{abstract}


\begin{highlights}
\item Hamiltonian-based architectures effectively model systems under nonholonomic constraints. 
\item The pseudo-Hamiltonian formulation enables the modeling of nonholonomic constraints from data.
\item A three network architecture simultaneously learns the Hamiltonian, constraints, and Lagrange multipliers of the system.
\end{highlights}


\end{frontmatter}


\section{Introduction}
\label{Introduction}
Data-driven approaches for the modeling of dynamical systems have seen a rise in popularity, with many works implementing methodologies such as sparse identification of nonlinear dynamics (SINDy) \cite{Sindy1,Sindy2,Sindy3}, dynamic mode decomposition (DMD) \cite{DMD1,DMD2,DMD3}, symbolic modeling \cite{Symbolic}, and machine learning \cite{PINN,HNN,GHNN,Nonholonomic-Robot}.
In the latter category, neural networks have been particularly successful, as they provide a versatile framework for modeling directly from
data that is well equipped to deal with high-dimensional problems \cite{weinan2021dawning,nakamura2021adaptive}.
An increasingly common approach is to incorporate physics priors to data-driven strategies \cite{Sindy3,DMD3,PINN}, improving modeling capability, computation time, and reducing model complexity. The incorporation of physics priors can be done explicitly, either embedding them in the architecture itself or in the loss function. \\


A family of methodologies with physics priors that has seen a growth in popularity are Hamiltonian-based models, and in particular, Hamiltonian neural networks (HNN)\cite{HNN}. HNNs explicitly use Hamiltonian mechanics as a prior in their formulation to enhance their modeling power. Greydanus et al. \cite{HNN} proposed to model Hamiltonian systems by learning the Hamiltonian function itself as the single output of a neural network, effectively encoding the dynamics of the system. Once the Hamiltonian is learned, autodifferentiation can be used to obtain its gradient, which together with Hamilton’s equations give the time derivative of the generalized coordinates and momenta. This formulation requires to learn a single function, the Hamiltonian, instead of the entire coordinate-derivative map. In principle, this reduces the number of parameters required to model the system and simplifies the function to be learnt by the network, potentially accelerating the training process and improving prediction accuracy. \\

The main disadvantage of the formulation proposed by Greydanus et al. is that it is limited to model classical Hamiltonian systems, which exclude any form of dissipation and systems under nonholonomic constraints. To improve on these limitations, multiple variations of the simple HNN architecture have been proposed. Sosanya and Greydanus \cite{DHNN} include dissipation in their formulation with Dissipative Hamiltonian Neural Networks (DHNN), learning a second scalar function that accounts for the nonconservative dynamics. Desai et al. \cite{Port-HNN} propose Port-Hamiltonian neural networks (Port-HNN), including both dissipation and external forces in the formulation. Course et al. \cite{GHNN} also extend on the simple HNN models with Generalized HNNs (GHNN), by embedding the generalized Hamiltonian decomposition \cite{GHF} on the architecture, extending the concept to a significantly more general family of systems. Another improvement is developed by Han et al. \cite{AHNN}, introducing Adaptable HNNs (AHNN), with more inputs in the first layer to account for variations in the system parameters.\\

Although Hamiltonian systems under nonholonomic constraints are in general not Hamiltonian, it is possible to formulate them in pseudo-Hamiltonian form \cite{NonholonomicFormulation}. 
There are multiple ways of equipping the nonholonomically constrained system with a bracket which is close to Poisson \cite{NonholonomicFormulation,edenrevisited}, one of them being Eden's bracket \cite{edenbracket}.
These brackets are said to not be fully Poisson as they do not necessarily satisfy the Jacobi identity.
This pseudo-Hamiltonian formulation opens the possibility to use the basic ideas of HNNs to model Hamiltonian systems under nonholonomic constraints. \\

The use of Hamiltonian-based architectures to model systems under nonholonomic constraints has been largely unexplored. Alltawaitan et al. \cite{Nonholonomic-Robot} use 
Hamiltonian neural ODE (HNODE) \cite{HNODE} to model the dynamics of an actuated wheeled robot under nonholonomic constraints. Nonetheless, their proposed architecture assumes total knowledge of the constraints of the system and is tailored to follow the constraints by construction. \\


The main contribution of this work is to propose a Hamiltonian-based neural network architecture capable of modeling systems under holonomic and nonholonomic constraints from data, while simultaneously learning the differential equations corresponding to the constraints and the forces they enact on the system. 

\section{Methodology}

\subsection{Hamiltonian mechanics and nonholonomic constraints}

Besides Newton's equations, there are other equivalent formulations of classical mechanics, which notably include Lagrangian and Hamiltonian mechanics. While Newtonian mechanics revolves around forces and their effect on systems, the other two define a  scalar function of the parameters of the system, from which the equations of motion are derived. These functions are referred to as the Lagrangian, which typically corresponds to the kinetic energy of the system minus its potential energy, and the Hamiltonian, which is obtained through the Legendre transformation of the Lagrangian. The Hamiltonian is usually the total energy of the system, and is expressed as a function of some set of generalized coordinates $\textbf{q}$ and the corresponding generalized momenta $\textbf{p}$ \cite{goldstein}. The Hamiltonian encodes all the information of the dynamics of conservative systems, which can be derived using Hamilton’s equations:

\begin{equation} \label{Heqs}
\dot{q_{i}}=\frac{\partial H}{\partial p_{i}} \quad ; \quad \dot{p_{i}}=-\frac{\partial H}{\partial q_{i}}
\end{equation}

Where $H$ is the Hamiltonian, $q_{i}$ and $p_{i}$ are the components of the vectors $\textbf{q}$ and $\textbf{p}$, and $\dot{q_{i}}$ and $\dot{p_{i}}$ symbolizes total derivative with respect to time. The vectors $\dot{\textbf{q}}$ and $\dot{\textbf{p}}$ are usually referred to as the phase velocity of the system.\\

This simple Hamiltonian formulation is limited to systems that are conservative, and is not equipped to deal with constraints in general. This limitation forbids its use for modeling systems under nonholonomic constraints, which are of special interest in engineering, robotics, mechanism design, path planning for autonomous vehicles, and system control among others. \\

A constraint is said to be nonholonomic if it is not possible to write it as a function of only the generalized coordinates $\textbf{q}$, requiring to be expressed as a function of  $\dot{\textbf{q}}$ or higher time derivatives. For the present work, we will consider linear nonholonomic constraints of the form:

\begin{equation} \label{NHC}
\textbf{A}\dot{\textbf{q}}=\textbf{b}
\end{equation}

\noindent where the system has $n$ degrees of freedom, $\dot{\textbf{q}}$ is a $n$-dimensional vector containing the time derivatives of the generalized coordinates, $\textbf{A}$ is a $m$ by $n$ matrix ($m$ being the number of constraints), and $\textbf{b}$ is a $m$-dimensional vector. Both $A$ and $\textbf{b}$ can be a function of $\textbf{q}$, but are assumed to be independent of $\dot{\textbf{q}}$. If $\textbf{b}=0$ the constraint is said to be catastatic, and in any other case the constraint is said to be acatastatic \cite{AnalyticalDynamics}. When eq. \ref{NHC} is non integrable, it necessarily corresponds to a nonholonomic constraint. \\

Note that holonomic constraints can also be expressed in the form shown in eq. \ref{NHC} when they are differentiated with respect to time, making this representation useful for both holonomic and nonholonomic constraints. Eq. \ref{NHC} can also be used to provide linear approximations to nonlinear nonholonomic constraints. This approach is particularly useful in low-velocity regimes or if the nonlinearities are small.\\

Eq. \ref{Heqs} can be extended to model the dynamics of a Hamiltonian system under constraints as in eq. \ref{NHC} \cite{NonholonomicFormulation}:

\begin{equation} \label{NHHeqs}
\dot{q_{i}}=\frac{\partial H}{\partial p_{i}} \quad ; \quad \dot{p_{i}}=-\frac{\partial H}{\partial q_{i}} + \sum_{j=1}^{m} A_{ij}^{T}\lambda_{j} 
\end{equation}

Where $A_{ij}^{T}$ are the components of $\textbf{A}^{T}$, and $\lambda_{j}$ are the Lagrange multipliers associated with each constraint. The summation term is a matrix multiplication of the form $\textbf{A}^{T}\boldsymbol{\lambda}$ where $\boldsymbol{\lambda}$ is a vector. This term represents the constraint forces on the system. \\

In order to analytically obtain the vector $\boldsymbol{\lambda}$, one can differentiate eq. \ref{NHC} with respect to time and write the resulting expression as a function of \textbf{q}, \textbf{p} and $\boldsymbol{\lambda}$ only. This can be done as long as the matrix components of $A$ are smooth, and it leads to a system of equations that can be solved to obtain the multipliers as a function of \textbf{q} and \textbf{p}:

\begin{equation} \label{ConstraintDerivative}
\dot{\textbf{A}}\dot{\textbf{q}}+\textbf{A}\ddot{\textbf{q}}=\dot{\textbf{b}}
\end{equation}

\noindent where $\dot{\textbf{A}}$, $\dot{\textbf{q}}$, $\textbf{A}$ and $\dot{\textbf{b}}$ are expressed as a function of $\textbf{q}$ and $\textbf{p}$, and $\ddot{\textbf{q}}$ is expressed as a function of $\textbf{q}$, $\textbf{p}$, and $\boldsymbol{\lambda}$.\\

For nonholonomic constraints we necessarily have that the eq. \ref{NHC} is nonintegrable. When this is the case, it can be seen as a second set of differential equations imposed on the system (the first one being eq. \ref{NHHeqs}), which is a major difference when compared to systems under holonomic constraints. Both sets of equations are usually coupled and nonlinear in the coordinates. Another distinguishing factor of systems under nonholonomic constraints is that they are not truly variational, as their equations of motion do not come directly from minimizing an action, but as a consequence of the Lagrange-d'Alambert principle \cite{blochnonholonomic}. \\

When a system is under a holonomic constraint, the dimensionality of its configuration space is reduced by one. This allows to write one coordinate as a function of the others, effectively turning it into an ignorable coordinate. This is not possible when a constraint is nonholonomic, as the conditions imposed on $\dot{\textbf{q}}$ lower the number of degrees of freedom of the system without restricting its configuration space.\\

An important property of holonomic and nonholonomic constraints as shown in eq. \ref{NHC} is that their representation is not unique, i.e., two different sets of quantities $\textbf{A},\textbf{b}$ and $\textbf{A}',\textbf{b}'$ may represent the same constraint. We show two relevant cases:\\

\begin{itemize}
    \item Linear combinations of the constraints: If a nonholonomic constraint is defined by a matrix $\textbf{A}$ and vector $\textbf{b}$, then a linear combination of said constraints defined by an invertible matrix $\textbf{M}$ is also a valid representation, as the matrix can be factored out when replacing the transformed quantities in eq. \ref{NHC}. The complete set of transformations is as follows:
    \begin{equation}\label{T1}
    \begin{split}
        \textbf{A} \rightarrow & \textbf{MA} \quad ; \quad \textbf{b} \rightarrow \textbf{M}\textbf{b} \\ & \boldsymbol{\lambda} \rightarrow 
        (\textbf{M}^{T})^{-1}\boldsymbol{\lambda}
    \end{split}
    \end{equation}

    Where $\textbf{M}$ is a matrix representing a linear combination of the constraints, and $(\textbf{M}^{T})^{-1}$ is the inverse of the transpose of $\textbf{M}$.\\
    
    \item Scaling: Multiplying the constraints by a generic nonzero function of \textbf{q} also leads to a valid representation, as the function can be factored out when replacing in eq. \ref{NHC}.  The complete set of transformations is as follows:
    \begin{equation}\label{T2}
    \begin{split}
        \textbf{A} \rightarrow  f&(\textbf{q})\textbf{A} \quad ; \quad \textbf{b} \rightarrow f(\textbf{q})\textbf{b} \\ & \boldsymbol{\lambda} \rightarrow 
        (f(\textbf{q}))^{-1}\boldsymbol{\lambda}
    \end{split}
    \end{equation}

    Where $f(\textbf{q})$ is a single scalar function of \textbf{q}, and $(f(\textbf{q}))^{-1}$ represents its multiplicative inverse.
\end{itemize}

\bigskip

It can be shown that the dynamics of the system remain unchanged under these transformations by using eqs. \ref{T1} and \ref{T2} in eq. \ref{NHHeqs} and recognizing that the constraint forces will be the same in all cases.\\

\subsection{Hamiltonian neural networks}

\begin{figure}[htbp]
    \centering
    \includegraphics[scale=0.15]{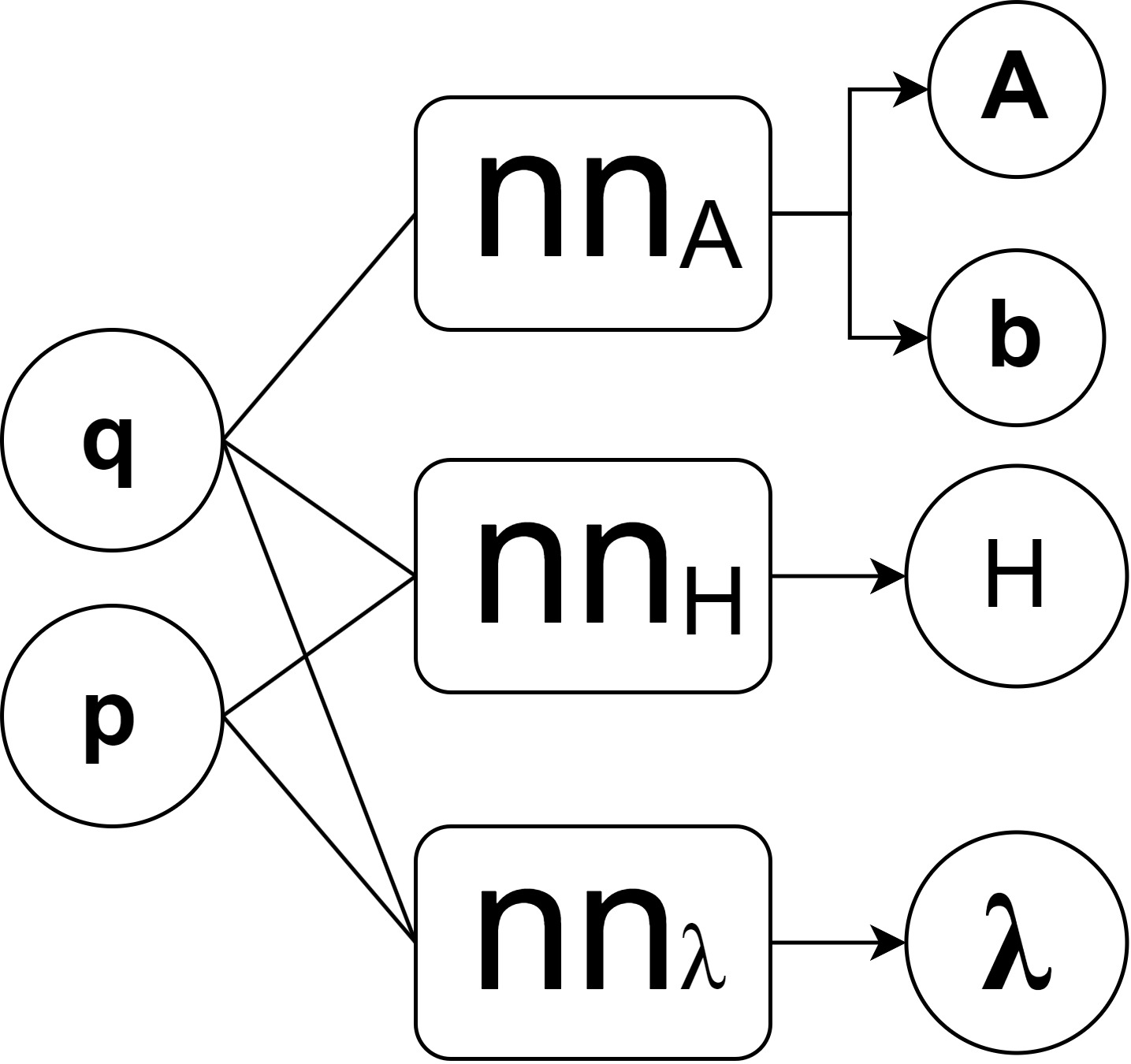}
    \caption{Proposed architecture to learn Hamiltonian systems under holonomic and nonholonomic constraints. Three independent networks are used to learn the Hamiltonian, the constraints through $\textbf{A}$ and $\textbf{b}$, and the multipliers in the vector $\boldsymbol{\lambda}$.} 
    \label{NHHNNdiagram}
\end{figure}

As previously stated, the formulation proposed by Greydanus et al. \cite{HNN} does not account for systems under nonholonomic constraints. To model Hamiltonian systems under constraints as in eq. \ref{NHC}, we propose a Hamiltonian based neural network architecture that uses eqs. \ref{NHHeqs} as a prior to model the system. The proposed architecture consists of three independent networks; the first network learns the Hamiltonian as a function of $\textbf{q}$ and $\textbf{p}$, the second network learns the components of the matrix $\textbf{A}$ (together with $\textbf{b}$ for acatastatic systems) as a function of $\textbf{q}$, while the third one learns the components of the vector $\boldsymbol{\lambda}$ as a function of $\textbf{q}$ and $\textbf{p}$, which correspond to the multipliers associated with the constraint forces. All the networks in this work are multilayer perceptrons. Figure \ref{NHHNNdiagram} shows the proposed architecture. \\

A single loss function is used to simultaneously train the three networks, explicitly incorporating equations \ref{NHC} and \ref{NHHeqs}:


\begin{equation} \label{NHHNNLoss}
    Loss=\sum_{i=1}^{N} \left( \dot{\textbf{q}}_{i}-\frac{\partial \hat{H}}{\partial \textbf{p}_{i}}\right)^{2}  + \left(\dot{\textbf{p}}_{i}+\frac{\partial \hat{H}}{\partial \textbf{q}_{i}} - \hat{\textbf{A}}^{T}\boldsymbol{\hat{\lambda}}\right)^{2} +
    \sum_{i=1}^{N} \left( \hat{\textbf{A}}\dot{\textbf{q}}_{i}-\boldsymbol{\hat{b}}\right)^{2}
\end{equation}

 \noindent where $\hat{H}$, $\hat{\textbf{A}}$,  $\boldsymbol{\hat{b}}$ and $\bm{\hat{\lambda}}$ are learned quantities and we sum over N points in the training set. The first sum in the loss function imposes eq. \ref{NHHeqs} to the three networks, while the second sum enforces the learned matrix $\hat{\textbf{A}}$ and the vector $\boldsymbol{\hat{b}}$ to be a valid representation of the constraints of the problem. \\

A relevant question is whether the three networks will learn exactly and only the quantity they are assigned to. In principle, the networks could learn in such a way that both $\dot{\textbf{q}}$ and $\dot{\textbf{p}}$ are properly modeled, but the terms $\hat{H}$ and $\hat{\textbf{A}}^{T}\boldsymbol{\hat{\lambda}}$ do not correspond to the actual Hamiltonian and constraint forces. In the supplementary material for this paper we derive a necessary condition to guarantee the uniqueness of the quantities learned by the networks.\\

Obtaining information about the Hamiltonian and constraints of the system is useful beyond the simple modeling of its behavior. A model for the Hamiltonian together with eq. \ref{Heqs} readily provides a model for the unconstrained system, while also being useful for energy-related applications such as noise prediction or the detection of dissipation. A model for the constraints provides essential information for applications in robotics and control, a well as being fundamental to model constraint forces.\\

We emphasize that the training of these networks does not require data on the Hamiltonian or the constraints; only data on $\textbf{q}$, $\textbf{p}$, $\dot{\textbf{q}}$ and $\dot{\textbf{p}}$ is needed.
If data is available for $\textbf{q}$ and $\dot{\textbf{q}}$ but not for $\textbf{p}$, numerical differentiation together with knowledge of the relation between $\dot{\textbf{q}}$, $\textbf{q}$ and  $\textbf{p}$ can be used to construct the necessary data.\\

The implementation shown in this paper is done entirely in PyTorch, and all networks are trained using the Adam optimizer. The dataset is divided into a training set and a validation set to monitor over fitting. The training is done in batches, which are constructed by randomly splitting the training set before each epoch. As the training of the networks progresses, the parameters corresponding to the lowest validation loss function are stored and later used to model the target system.

\section{Numerical experiments and results}

We choose two canonical examples of nonholonomically constrained systems to asses the performance of the proposed architecture; A disk rolling over a flat table, and a ball rolling on a spinning table. A third example, consisting of a nonholonomically constrained particle in three dimensions, is included in the supplementary material of this paper.

\subsection{Rolling disk}

\begin{figure}[htbp]
    \centering
    \includegraphics[width=\textwidth]{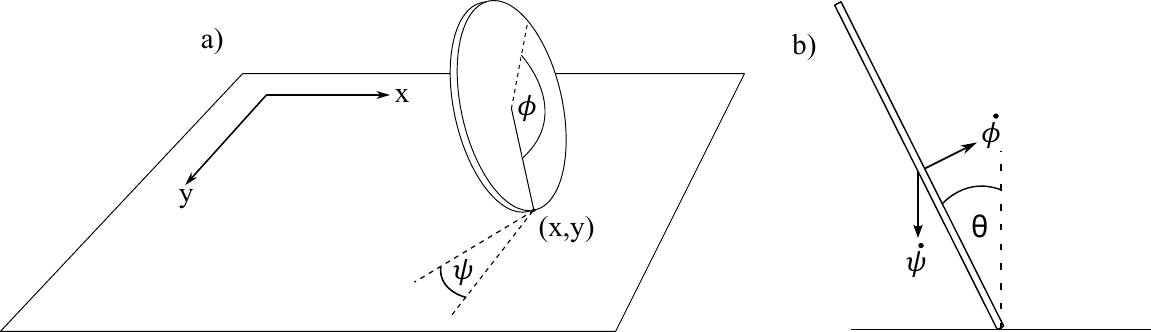}
    \caption{a) The coordinates $x$ and $y$ indicate the position of the point of contact between the disk and the surface, $\phi$ represents the rolling angle of the disk about its symmetry axis, and $\psi$ is the angle of rotation with respect to the vertical axis. b) $\theta$ is the leaning angle of the disk from its fully vertical position.} 
    \label{Disk}
\end{figure}

The rolling disk is a classical example of a Hamiltonian system under nonholonomic constraints. The system consists of a disk rolling without slipping on a flat surface, condition that needs to be expressed as a pair of nonholonomic constraints. Figure \ref{Disk} shows the coordinates used to describe the motion of the disk; $x$ and $y$ describe the position of the contact point between the disk and the surface, $\phi$ describes the angle of rotation about its axis of symmetry, $\psi$ describes its angle of rotation about the vertical axis in the fixed frame, and $\theta$ describes the leaning angle of the disk. \\

The Hamiltonian of the system is 
\begin{equation} \label{H2}
\begin{split}
    H=&\frac{p_{x}^{2}+p_{y}^{2}}{2m}+\frac{p_{\phi}^{2}}{2I_{xx}}+
    \frac{(p_{\psi}+p_{\phi}\sin(\theta))^{2}}{\cos^{2}(\theta)I_{xx}}
    \\&+\frac{p_{\theta}^{2}}{2(mR^{2}\sin^{2}(\theta)+\frac{I_{xx}}{2})}
    +mgR\cos(\theta)
\end{split}
\end{equation}

\vspace{2mm}

\noindent where $I_{xx}$ is the moment of inertia of the disk with respect to its main axis, $m$ and $R$ are the mass and radius of the disk, and $g$ is the acceleration due to gravity. The generalized momenta of the system take the form

\begin{equation} \label{p2}
\begin{split}
&p_{x}=m\dot{x} \quad ; \quad
p_{y}=m\dot{y} \quad ; \quad \\
&p_{\theta}=\dot{\theta}(mR^{2}\sin^{2}(\theta)+\frac{I_{xx}}{2})\quad ; \quad\\
& p_{\psi}=I_{xx}\dot{\phi}-I_{xx}\sin(\theta)\dot{\psi} \quad ; \quad \\
&p_{\phi}=I_{xx}\dot{\psi}\left(1-\frac{\cos^{2}(\theta)}{2}\right)
-I_{xx}\dot{\phi}\sin{\theta}
\end{split}
\end{equation}

\bigskip

Two nonholonomic constraints are needed to impose the no-slip condition to the system, and they take the form

\begin{equation} \label{c2}
\begin{split}
\dot{x}=&-R\sin(\psi)(\dot{\phi}-\dot{\psi}\sin(\theta))\\
&-R\dot{\theta}\cos(\psi)\cos(\theta)\\
\dot{y}=&R\cos(\psi)(\dot{\phi}-\dot{\psi}\sin(\theta))\\
&-R\dot{\theta}\sin(\psi)\cos(\theta)
\end{split}
\end{equation}

\bigskip

The phase space of the system has ten dimensions and there are two constraints, therefore the networks associated with the Hamiltonian and the multipliers take ten inputs and give one and two outputs respectively, while the network assigned to learn the $\textbf{A}$ matrix has five inputs and ten outputs. Table \ref{TableRD} summarizes the parameters used for training and the networks used in each case for the rolling disk. The learning rates for the constraint and multiplier networks were reduced to $2e-5$ after $1.5e5$ epochs, and to $1e-5$ after $3e5$ epochs, as it was found empirically to facilitate convergence.\\

\begin{table}[htbp]
\centering
\caption{Parameters used for the rolling disk experiment.}
\label{TableRD}
\begin{tabular}{|l|ccc|}
\hline
\multicolumn{1}{|c|}{\textbf{Parameter}} & \multicolumn{1}{c|}{\textbf{$H$}} & \multicolumn{1}{c|}{\textbf{$\bm{\lambda}$}} & \textbf{$\textbf{A}$} \\ \hline
Hidden layers                            & \multicolumn{1}{c|}{2}                                    & \multicolumn{1}{c|}{2}                                          & 2                                    \\ \hline
Neurons per   layer                      & \multicolumn{1}{c|}{200}                                  & \multicolumn{1}{c|}{300}                                        & 400                                  \\ \hline
Activation   function                    & \multicolumn{1}{c|}{$Tanh$}                                 & \multicolumn{1}{c|}{$Tanh$}                                       & $Tanh$                                 \\ \hline
Learning rate                            & \multicolumn{1}{c|}{5e-5}                                 & \multicolumn{1}{c|}{5e-5*}                                       & 5e-5*                                 \\ \hline
Epochs                                   & \multicolumn{3}{c|}{5e5}                                                                                                                                         \\ \hline
Training set                             & \multicolumn{3}{c|}{4.8e4 points}                                                                                                                                  \\ \hline
Validation set                           & \multicolumn{3}{c|}{3e3 points}                                                                                                                                    \\ \hline
Batches                                  & \multicolumn{3}{c|}{48}                                                                                                                                            \\ \hline
\end{tabular}
\end{table}

To test the architecture, we use a sample trajectory with initial conditions $[x,y,\theta,\psi,\phi,p_{x},p_{y},p_{\theta},p_{\psi},p_{\phi}]=[2,0,\pi/8,0,0,0,0,0,p_{\psi\circ},p_{\phi\circ}]$ in phase space where $p_{\psi\circ}$ and $p_{\phi\circ}$ are the momenta corresponding to 

\begin{equation}
\dot{\psi}=\sqrt{\frac{4g\tan(\pi/8)}{12+r\sin(\pi/8)}} \quad ; \quad \dot{\phi}=\frac{2\dot{\psi}}{r}+\sin(\pi/8)\dot{\psi}
\end{equation}

The trajectory is chosen as it corresponds to a case of particular interest in which the disk rolls in a stable circle. The dataset is generated artificially, sampling uniformly distributed points in the neighborhood of the circular trajectory, as shown in Table \ref{RDranges}. Note that the range of $[0,2\pi]$ chosen for $\psi$ and $\phi$ is enough to cover all possible values of these variables as we know they are angles, and we can transform whatever input is given to the network to the $[0,2\pi]$ range.\\

\begin{table}[htbp]
\centering
\caption{Ranges of values for generated points.}
\label{RDranges}
\begin{tabular}{|c|c|c|}
\hline
\multicolumn{1}{|l|}{\textbf{Coordinate}} & \multicolumn{1}{l|}{\textbf{Lower limit}} & \multicolumn{1}{l|}{\textbf{Upper limit}} \\ \hline
$x$                                       & $-2.5$                                      & $2.5$                                       \\ \hline
$y$                                       & $-2.5$                                      & $2.5$                                       \\ \hline
$\theta$                                  & $0.6(\pi/8)$                              & $1.4(\pi/8)$                              \\ \hline
$\psi$                                    & $0$                                         & $2\pi$                                    \\ \hline
$\phi$                                    & $0$                                         & $2\pi$                                    \\ \hline
$p_{x}$                                   & $-3$                                        & $3$                                         \\ \hline
$p_{y}$                                   & $-3$                                        & $3$                                         \\ \hline
$p_{\theta}$                              & $-0.4$                                      & $0.4$                                       \\ \hline
$p_{\psi}$                                 & $-0.273$                                    & $-0.117$                                    \\ \hline
$p_{\phi}$                                 & $0.684$                                     & $1.596$                                     \\ \hline
\end{tabular}
\end{table}

\begin{figure}[htbp]
    \centering
    \includegraphics[width=0.64\textwidth]{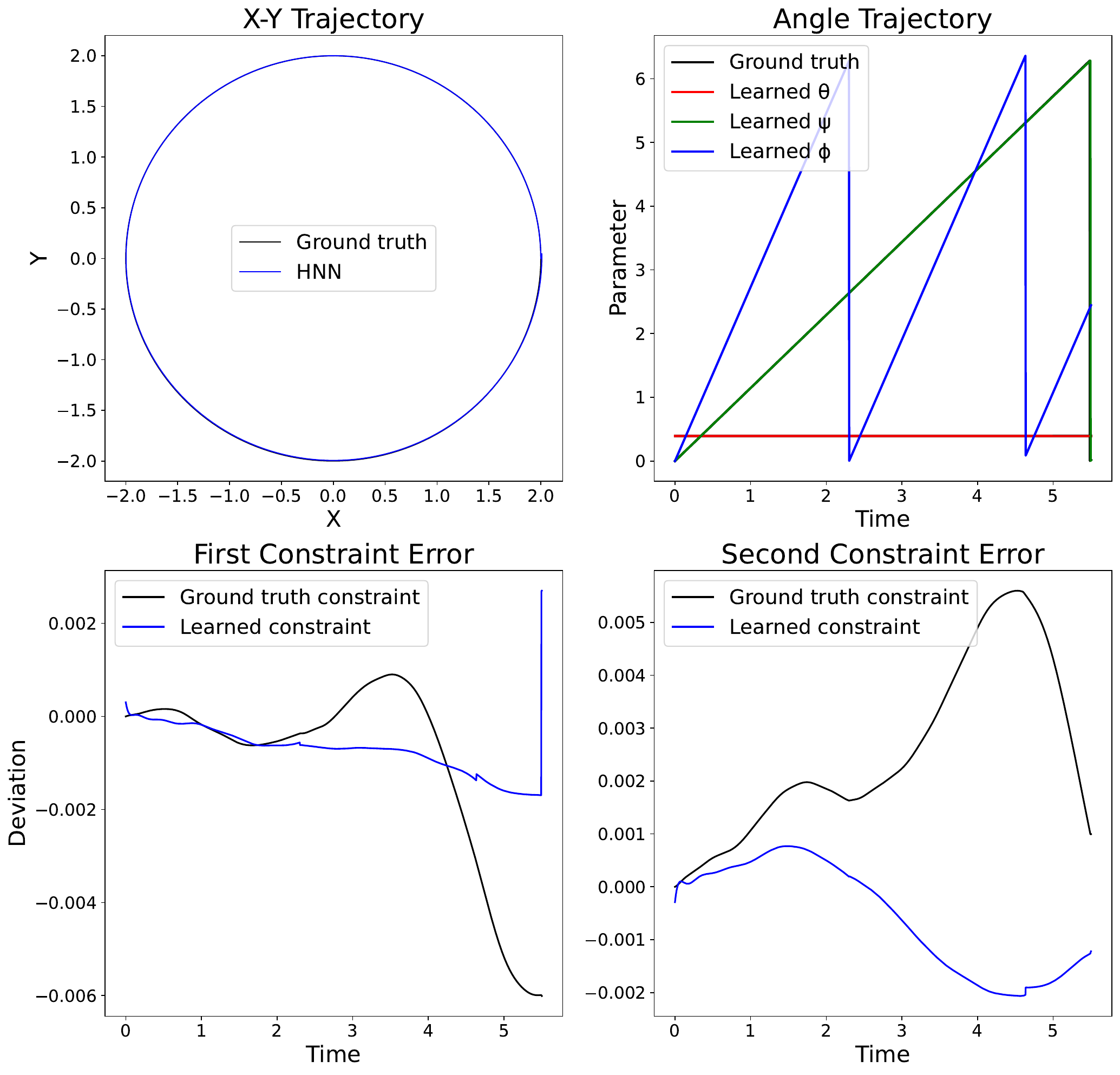}
    \caption{Trajectories and deviations for the rolling disk system.} 
    \label{RD_Traj}
\end{figure}

Top row of Figure \ref{RD_Traj} shows the evolution of the position and orientation of the disk along the learned and ground truth trajectory, showing the angles in the $[0,2\pi]$ range.\\

In order to asses how precisely the constraints hold, we can define the deviation from the analytical constraints as $\textbf{A} \dot{\boldsymbol{q}}-\boldsymbol{b}$, and $\hat{\textbf{A}}\dot{\textbf{q}}-\boldsymbol{\hat{b}}$ for the learned constraints, where $\dot{\textbf{q}}$ corresponds to the velocities of the modeled (not ground truth) trajectory. Bottom row of Figure \ref{RD_Traj} shows the the deviation from the constraints of the system. It can be seen that the ground truth trajectory is recovered in high precision both in the $x-y$ plane and in the angle variables. Both the learned and ground truth constraints present small deviations, with the former being closer to zero. This is probably due to the fact that the loss function in eq. \ref{NHHNNLoss} has a term that explicitly penalizes nonzero values for $\hat{\textbf{A}}\dot{\textbf{q}}-\boldsymbol{\hat{b}}$, while $\textbf{A}\dot{\textbf{q}}-\boldsymbol{b}$ is only minimized implicitly (i.e. it will be minimized as the correct trajectory is learned, but the term is not explicitly included in the loss function). The abrupt increase in deviation for the first learned constraint towards the end of the trajectory is due to the rapid change in $\psi$ as it is normalized to the $[0,2\pi]$ range.\\ 

\subsection{Ball on spinning table}

For the last experiment, we choose to work with a system consisting of a ball rolling without slipping over a spinning table. The table/surface spins with constant angular velocity $\Omega$ and is considered to be infinite in size. An interesting aspect of the system is that, unlike in the previous cases, its constraints are acatastatic, meaning $\textbf{b}$ in eq. \ref{NHC} is no longer zero, and as a consequence the energy of the ball will not be a constant of motion. \\

\begin{figure}[htb!]
    \centering
    \includegraphics[width=0.7\textwidth]{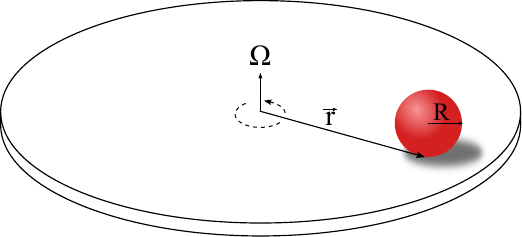}
    \caption{The position of the ball with respect to the center of the table is given by the vector $\textbf{r}=(x,y)$. The table spins at constant angular velocity $\Omega$ and is considered to be infinite in size.} 
    \label{BoT}
\end{figure}

In order to formulate the system from a Hamiltonian viewpoint, we need to account not only for the position of the ball, but also its orientation. To represent the orientation of the ball we choose to use the Euler-Rodrigues parameters, which provide a useful representation for all possible rotations over a rigid body and do not present gimbal locking. In order to use these rotations to keep track of the orientation of the ball, we can think of a reference orientation of the ball that is rotated to achieve the new states, effectively mapping from rotations to orientations. The Euler-Rodrigues parameters are related to the angle-axis parameter as \cite{MultibodyDynamics}

\begin{equation} \label{E-R}
\begin{split}
&e_{0}=\cos\left(\frac{\alpha}{2}\right)\\
&e_{1}=k_{1}\sin\left(\frac{\alpha}{2}\right)\\
&e_{2}=k_{2}\sin\left(\frac{\alpha}{2}\right)\\
&e_{3}=k_{3}\sin\left(\frac{\alpha}{2}\right)
\end{split}
\end{equation}

\noindent where $e_{i}$ are the Euler-Rodrigues parameters, $k_{i}$ are the components of the axis unit vector, and $\alpha$ is the angle of rotation. As for the displacement of the ball, we simply use $x$ and $y$ to represent the position of its contact point with respect to the center of the table. Figure \ref{BoT} illustrates the system.\\

The Hamiltonian of the system is 
\vspace{10px}
\begin{equation} \label{H3}
    H=\frac{p_{x}^{2}+p_{y}^{2}}{2m}+\frac{p^{2}_{e0}+p^{2}_{e1}+p^{2}_{e2}+p^{2}_{e3}}{8I}
\end{equation}

\vspace{10px}
where $I$ is the moment of inertia of the ball and $m$ its mass. As in the previous cases, from eq. \ref{Heqs} we can derive the generalized momenta, which take the form

\begin{equation} \label{p3}
\begin{split}
&p_{x}=m\dot{x} \quad ; \quad
p_{y}=m\dot{y} \quad ; \quad \\
&p_{e0}=4I\dot{e_{0}} \quad ; \quad
p_{e1}=4I\dot{e_{1}} \quad ; \quad \\
&p_{e2}=4I\dot{e_{2}} \quad ; \quad
p_{e3}=4I\dot{e_{3}} \quad ; \quad \\
\end{split}
\end{equation}

The nonholonomic constraints associated with the no-slip condition are
\vspace{2mm}
\begin{equation} \label{c3}
\begin{split}
&\dot{x}+2R(e_{2}\dot{e_{0}}-e_{3}\dot{e_{1}}-e_{0}\dot{e_{2}}+e_{1}\dot{e_{3}})=-\Omega  y\\
&\dot{y}+2R(-e_{1}\dot{e_{0}}+e_{0}\dot{e_{1}}-e_{3}\dot{e_{2}}+e_{2}\dot{e_{3}})=\Omega  x
\end{split}
\end{equation}

\vspace{2mm}
\noindent where $R$ is the radius of the ball and $\Omega$ is the angular velocity of the table.\\

The system has twelve dimensions in its phase space and there are two constraints, therefore the networks associated with the Hamiltonian and the multipliers take twelve inputs and give one and two outputs respectively, while the network assigned to learn $\textbf{A}$ and $\textbf{b}$ has six inputs and fourteen outputs. Table \ref{TableBoT} summarizes the parameters used for training in each case for the ball on the spinning table. Table \ref{RDIC} shows the initial conditions chosen to test the architecture. These values are chosen to keep the magnitudes of all variables reasonably bounded along the trajectory.\\

\begin{table}[htbp]
\centering
\caption{Parameters used for the ball on a spinning table experiment.}
\label{TableBoT}
\begin{tabular}{|l|ccc|}
\hline
\multicolumn{1}{|c|}{\textbf{Parameter}} & \multicolumn{1}{c|}{{$H$}} & \multicolumn{1}{c|}{$\bm{\lambda}$} & {$\textbf{A}$} \\ \hline
Hidden layers                            & \multicolumn{1}{c|}{2}                                    & \multicolumn{1}{c|}{2}                                          & 2                                    \\ \hline
Neurons per layer                        & \multicolumn{1}{c|}{200}                                  & \multicolumn{1}{c|}{200}                                        & 50                                   \\ \hline
Activation function                      & \multicolumn{1}{c|}{$Tanh$}                                 & \multicolumn{1}{c|}{$Tanh$}                                       & $Tanh$                                 \\ \hline
Learning rate                            & \multicolumn{1}{c|}{5e-5}                                 & \multicolumn{1}{c|}{5e-5}                                       & 5e-5                                 \\ \hline
Epochs                                   & \multicolumn{3}{c|}{3e5}                                                                                                                                           \\ \hline
Training set                             & \multicolumn{3}{c|}{1.2e4 points}                                                                                                                                  \\ \hline
Validation set                           & \multicolumn{3}{c|}{3e3 points}                                                                                                                                    \\ \hline
Batches                                  & \multicolumn{3}{c|}{12}                                                                                                                                            \\ \hline
\end{tabular}
\end{table}

\begin{table}[htbp]
\centering
\caption{Initial conditions for the ball on a spinning table experiment.}
\label{RDIC}
\begin{tabular}{
>{\columncolor[HTML]{FFFFFF}}c 
>{\columncolor[HTML]{FFFFFF}}c 
>{\columncolor[HTML]{FFFFFF}}c 
>{\columncolor[HTML]{FFFFFF}}c 
>{\columncolor[HTML]{FFFFFF}}c 
>{\columncolor[HTML]{FFFFFF}}c }
\hline
\multicolumn{1}{|c|}{\cellcolor[HTML]{FFFFFF}x}        & \multicolumn{1}{c|}{\cellcolor[HTML]{FFFFFF}y}        & \multicolumn{1}{c|}{\cellcolor[HTML]{FFFFFF}$e_{0}$}        & \multicolumn{1}{c|}{\cellcolor[HTML]{FFFFFF}$e_{1}$}        & \multicolumn{1}{c|}{\cellcolor[HTML]{FFFFFF}$e_{2}$}        & \multicolumn{1}{c|}{\cellcolor[HTML]{FFFFFF}$e_{3}$}        \\ \hline
\multicolumn{1}{|c|}{\cellcolor[HTML]{FFFFFF}0.0264}   & \multicolumn{1}{c|}{\cellcolor[HTML]{FFFFFF}0.1283}   & \multicolumn{1}{c|}{\cellcolor[HTML]{FFFFFF}0.2644}          & \multicolumn{1}{c|}{\cellcolor[HTML]{FFFFFF}0.962}           & \multicolumn{1}{c|}{\cellcolor[HTML]{FFFFFF}-0.0423}         & \multicolumn{1}{c|}{\cellcolor[HTML]{FFFFFF}-0.0535}         \\ \hline
\multicolumn{1}{l}{\cellcolor[HTML]{FFFFFF}}           & \multicolumn{1}{l}{\cellcolor[HTML]{FFFFFF}}          & \multicolumn{1}{l}{\cellcolor[HTML]{FFFFFF}}                 & \multicolumn{1}{l}{\cellcolor[HTML]{FFFFFF}}                 & \multicolumn{1}{l}{\cellcolor[HTML]{FFFFFF}}                 & \multicolumn{1}{l}{\cellcolor[HTML]{FFFFFF}}                 \\ \hline
\multicolumn{1}{|c|}{\cellcolor[HTML]{FFFFFF}$p_{x}$} & \multicolumn{1}{c|}{\cellcolor[HTML]{FFFFFF}$p_{y}$} & \multicolumn{1}{c|}{\cellcolor[HTML]{FFFFFF}$p_{e_{0}}$} & \multicolumn{1}{c|}{\cellcolor[HTML]{FFFFFF}$p_{e_{1}}$} & \multicolumn{1}{c|}{\cellcolor[HTML]{FFFFFF}$p_{e_{2}}$} & \multicolumn{1}{c|}{\cellcolor[HTML]{FFFFFF}$p_{e_{3}}$} \\ \hline
\multicolumn{1}{|c|}{\cellcolor[HTML]{FFFFFF}-0.2957}  & \multicolumn{1}{c|}{\cellcolor[HTML]{FFFFFF}-0.1952}  & \multicolumn{1}{c|}{\cellcolor[HTML]{FFFFFF}-0.1993}         & \multicolumn{1}{c|}{\cellcolor[HTML]{FFFFFF}0.0486}          & \multicolumn{1}{c|}{\cellcolor[HTML]{FFFFFF}-0.1246}         & \multicolumn{1}{c|}{\cellcolor[HTML]{FFFFFF}-0.0131}         \\ \hline
\end{tabular}
\end{table}


\subsubsection{Solid sphere}



\begin{figure}[htb!]
    \centering
    \includegraphics[width=0.61\textwidth]{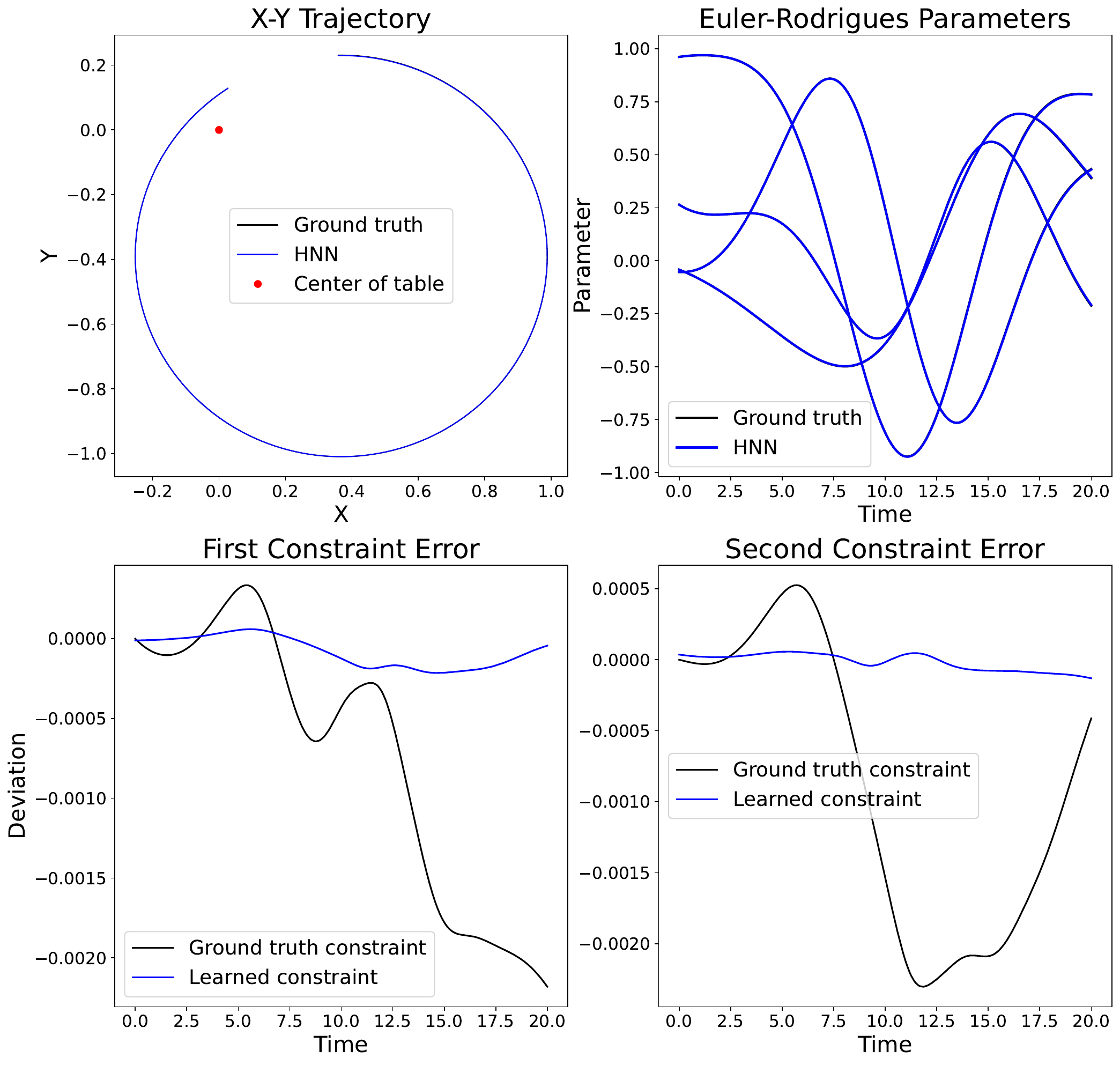}
    \caption{Trajectories and deviations for the solid ball on spinning table. } 
    \label{BoT_Traj}
\end{figure}

We first consider the case $I=\frac{2mR^{2}}{5}$, corresponding to the case of a solid sphere.
Top row of Figure \ref{BoT_Traj} shows the evolution of the position and Euler-Rodrigues parameters of the ball along the learned and ground truth trajectory. We show the deviations from the constraints in the bottom row of Figure \ref{BoT_Traj}, defined as in the previous experiment. Similarly to the rolling disk system, the ground truth constraints and specially the learned constraints stay close to zero, which implies that the constrained behavior of the system is effectively recovered.\\ 

\subsubsection{Hollow sphere}



\begin{figure}[htb!]
    \centering
    \includegraphics[width=0.61\textwidth]{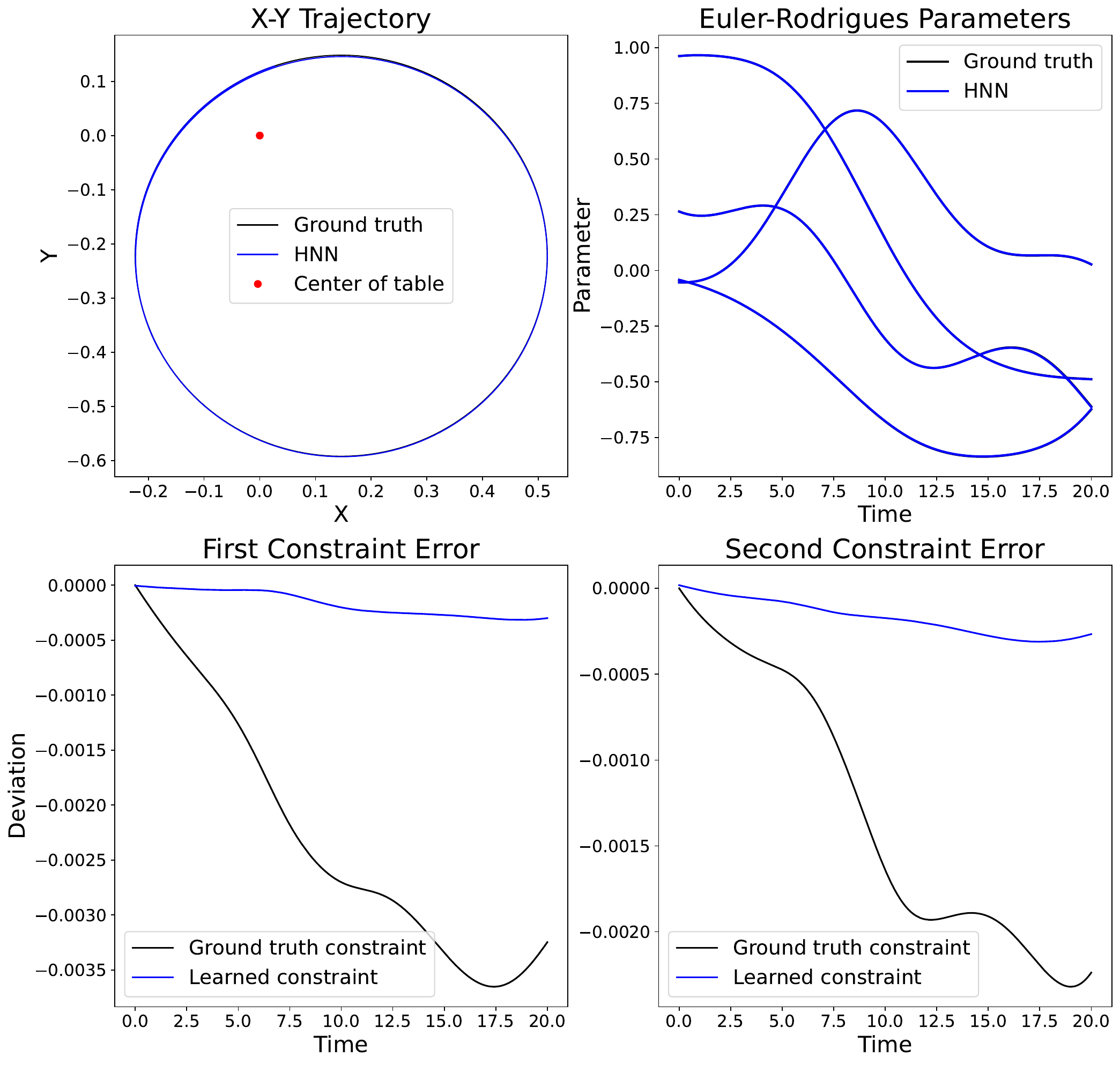}
    \caption{Trajectories and deviations for the hollow ball on spinning table.} 
    \label{HBoT_Traj}
\end{figure}

We now consider the case $I=\frac{2mR^{2}}{3}$, corresponding to the case of a hollow sphere.
Top row of Figure \ref{HBoT_Traj} shows the evolution of the position and Euler-Rodrigues parameters, and the deviations from the constraints are shown in the bottom row of Figure \ref{HBoT_Traj}. Similarly the solid sphere case, Figure \ref{HBoT_Traj} shows that the system is modeled with high precision, effectively recovering the main behaviors and parameters. \\

\subsection{Performance in the presence of noise}

The experiments in the previous subsections were repeated with a noisy dataset to assess the performance of the proposed architecture under more realistic conditions. A multiplicative noise model is used, where every variable $x$ in the data matrix is independently modified as:

\begin{equation}\label{noise}
    x \rightarrow x(1+\mathcal{N}(0,\sigma^{2}))
\end{equation}

\noindent where $\mathcal{N}(0,\sigma^{2})$ is a number sampled from a normal distribution with zero mean and a standard deviation of $\sigma=0.02$. We note that, in order to asses overfitting due to noise more effectively, we use eq. \ref{noise} on the training data but not to the validation set. In order to emulate the repeated measurements that are often taken in real experiments to deal with noise, ten iterations of the noisy dataset are computed (all from the same ``clean'' original dataset), and their average is used to train the networks.\\

The networks are trained exactly as in the case without noise for the ball on spinning table, using the parameters in tables  \ref{TableBoT}. The rolling disk case is trained with similar parameters as shown in \ref{TableRD}, but the number of epochs is reduced to $5e4$ as the noisy dataset causes over fitting early in the training process. As with the previous experiments, the ground truth and learned constraint errors are defined as $\textbf{A} \dot{\boldsymbol{q}}-\boldsymbol{b}$ and $\hat{\textbf{A}}\dot{\textbf{q}}-\boldsymbol{\hat{b}}$ where $\boldsymbol{q}$ corresponds to the modeled trajectory.

\subsubsection{Rolling disk: Noisy dataset}



\begin{figure}[htb!]
    \centering
    \includegraphics[width=0.64\textwidth]{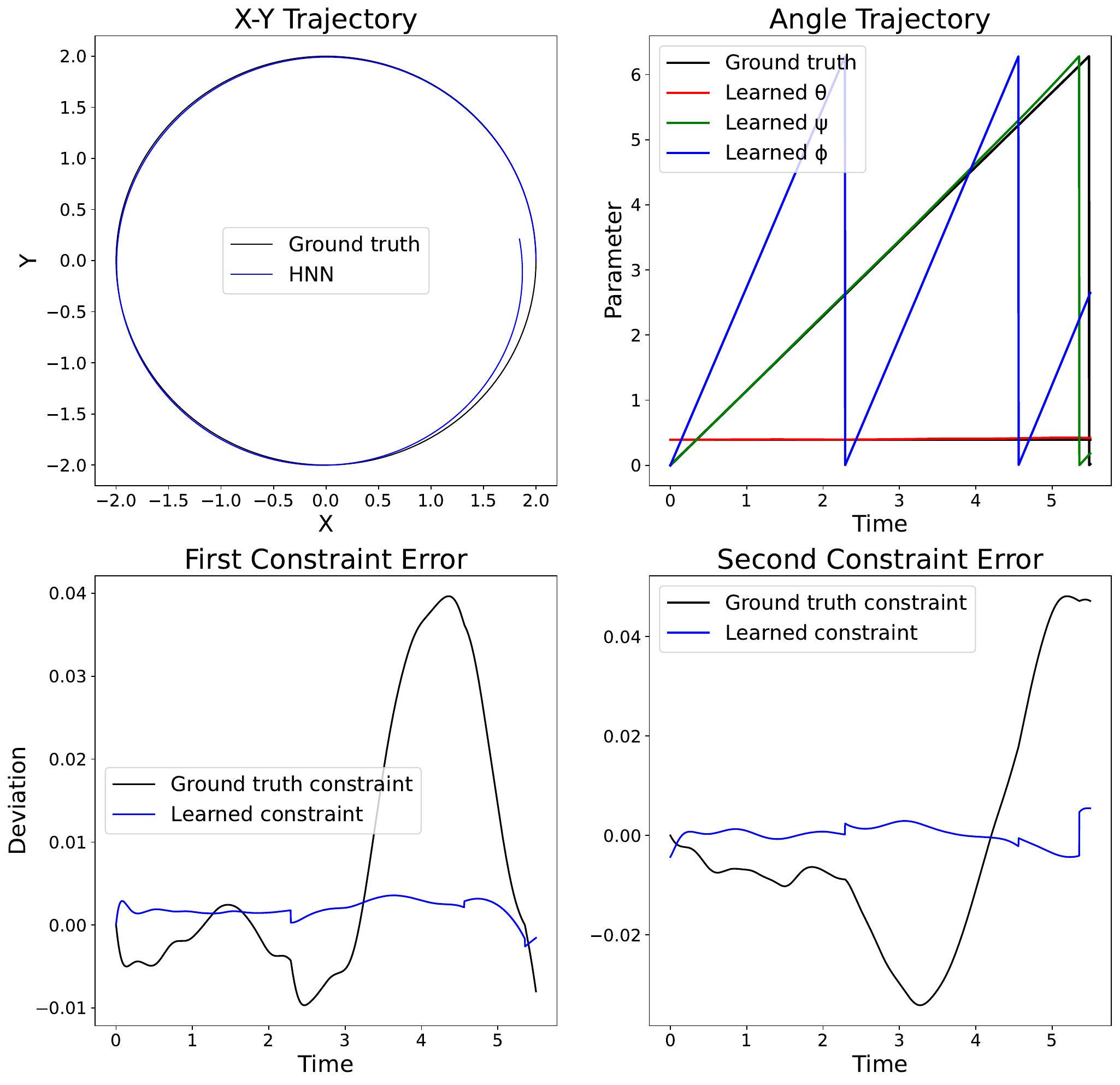}
    \caption{Trajectories and deviations for the rolling disk system in the presence of noise. } 
    \label{RD_Traj_noise}
\end{figure}

Top row of Figure \ref{RD_Traj_noise} shows the $x-y$ trajectory of the rolling disk and the evolution of its orientation over time. Bottom row shows deviations from the analytical and learned constraints. It can be seen how the trajectory is initially recovered with high precision, but after some time the larger errors due to noise accumulate and lead to a clear divergence when compared to the ground truth. Similarly to the case without noise, the learned constraint deviations are lower than the ground truth ones, but now their values are around one order of magnitude larger.\\


\subsubsection{Solid ball on table: Noisy dataset}



\begin{figure}[htb!]
    \centering
    \includegraphics[width=0.64\textwidth]{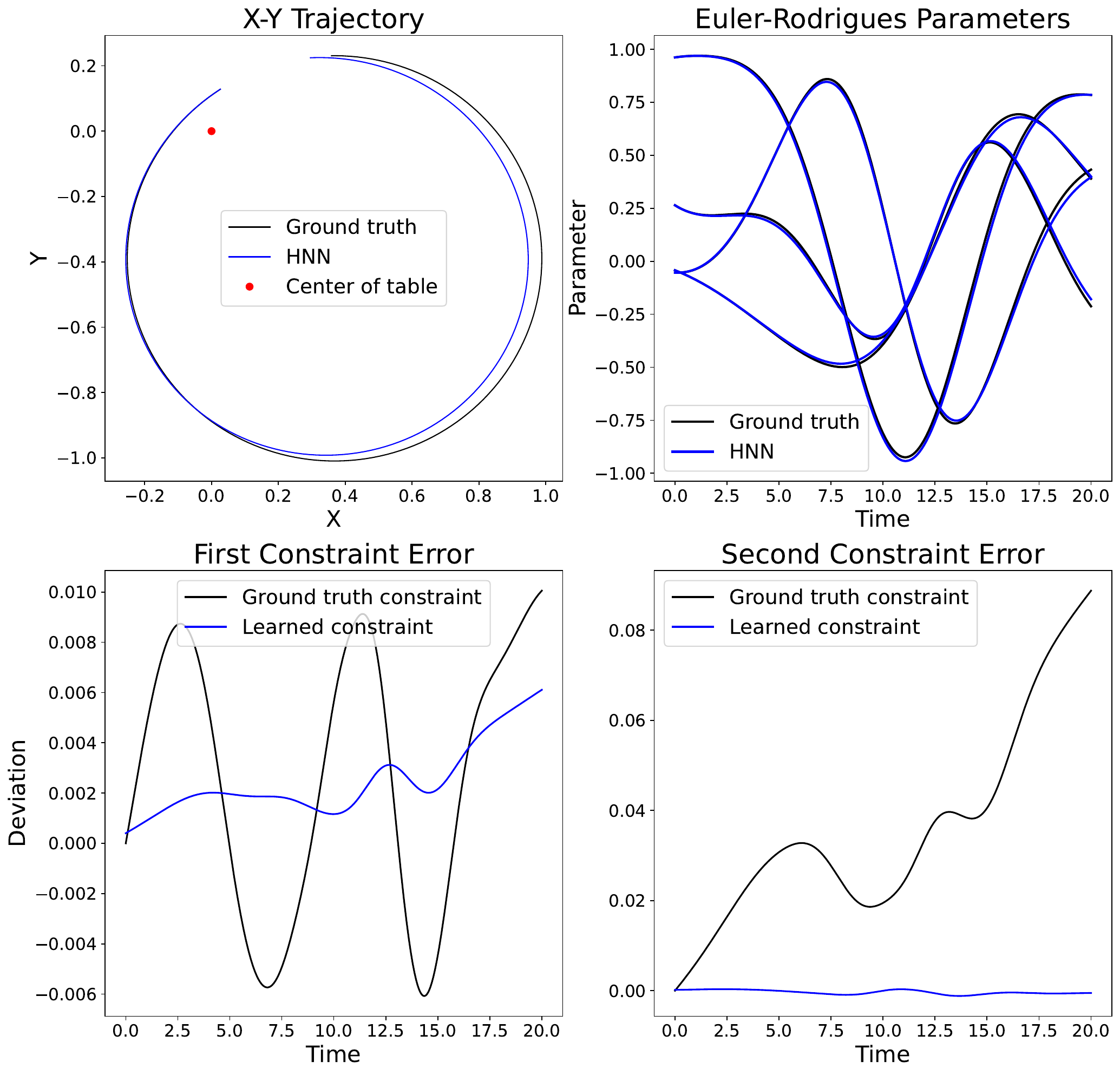}
    \caption{Trajectories and deviations for the solid ball on spinning table with noise. } 
    \label{FBoT_Traj_noise}
\end{figure}

Top row of Figure \ref{FBoT_Traj_noise} shows the $x-y$ trajectory of the ball and the evolution of its orientation over time. Bottom row shows deviations from the analytical and learned constraints. Similarly to the rolling disk case, it can be seen that the trajectories are initially recovered but deviate from the ground truth with time, and that the constraint deviations are around one order of magnitude larger when compared to the case without noise. 

\subsubsection{Hollow ball on table: Noisy dataset}

\begin{figure}[htb!]
    \centering
    \includegraphics[width=0.64\textwidth]{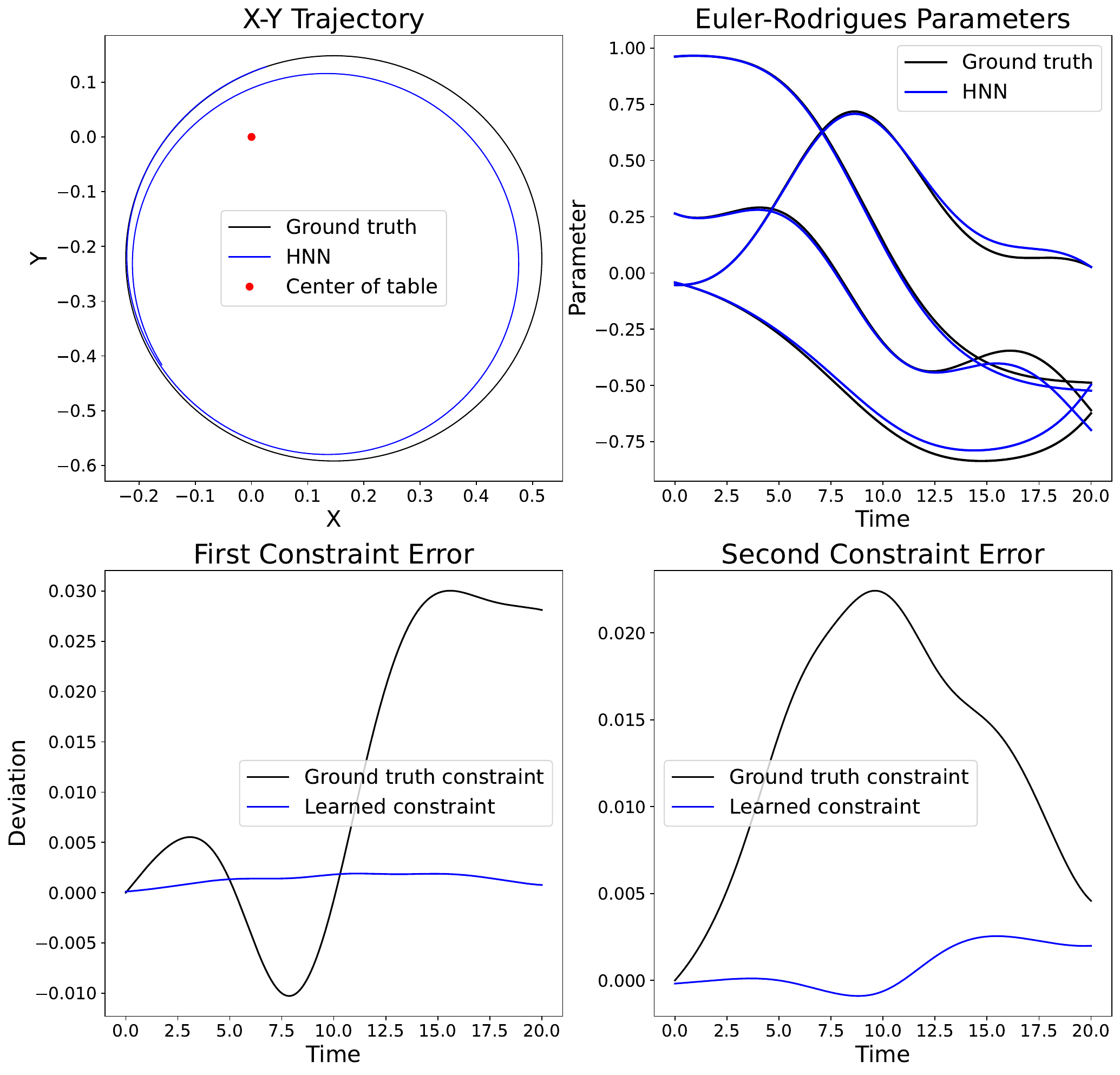}
    \caption{Results for the hollow ball on spinning table with noise. } 
    \label{HBoT_Traj_noise}
\end{figure}

Top row of Figure \ref{HBoT_Traj_noise} shows the $x-y$ trajectory of the ball and the evolution of its orientation over time. Bottom row shows deviations from the analytical and learned constraints. Results are similar to the solid sphere case with noise, with trajectories that are initially precise but deviate over time, and higher constraint deviations when compared to the case without noise.

\subsection{General results}

In this subsection we summarize results corresponding to the training time, loss functions, stability of the predicted trajectories over time, and precision of the learned Hamiltonian and constraint forces.\\

To asses the precision of the learned Hamiltonian, the ground truth energy of the validation set of all systems is compared with the predicted Hamiltonian using the mean absolute error (MAE). To deal with the fact that energy is unique only up to a constant, the means of the predicted and ground truth sets are displaced to be equal before computing the MAE. \\

Similarly, the MAE is used to evaluate the average error of the modeled constraint forces, where we define the error at each point as the magnitude of the vector $\textbf{A}^{T}\boldsymbol{\lambda}-\hat{\textbf{A}}^{T}\hat{\boldsymbol{\lambda}}$, the first term corresponding to the analytical quantities and the second to modeled quantities. The MAE is then normalized by the mean value of the analytical force $\textbf{A}^{T}\boldsymbol{\lambda}$.\\


To asses the stability of the predictions, Lyapunov exponents are computed for the deviation of the modeled trajectories with respect to the ground truth. The deviation is defined as the vector containing the magnitude of the phase space error at each point. The initial $30\%$ of this vector is removed as its slope in logarithmic scale seems to stabilize after this point. The resulting vector is normalized to start with a value of $1$.
The inverse of the Lyapunov exponents are also shown, corresponding to the characteristic divergence times.\\

Tables \ref{GRNoNoise} and \ref{GRWNoise} show the results for the experiments without noise and with noise, respectively. In both cases the system that required the most time to be modeled is the rolling disk, which implies that the complexity of the modeling task depends on more than just the dimensionality of the system. In the case without noise, both the energy and the forces are learned with high precision, which speaks of an effective recovery of the Hamiltonian, constraints, and multipliers of the system. Although the modeling error of the energy increases significantly in the presence of noise, this does not seem to be the case for the constraint forces, which maintain a low normalized deviation. This may suggest high robustness of the learned constraints and multipliers to imperfections in the training set. The Lyapunov exponents (and therefore the characteristic divergence times) are shown to be of the same order of magnitude across experiments, which speaks of the stability of the proposed architecture.

\begin{table}[htb!]
\centering
\caption{Training times, energy mean absolute error (MAE), Lyapunov exponent and characteristic times for the experiments without noise.}
\label{GRNoNoise}
\scalebox{0.85}{
\begin{tabular}{|
>{\columncolor[HTML]{FFFFFF}}c |
>{\columncolor[HTML]{FFFFFF}}c |
>{\columncolor[HTML]{FFFFFF}}c |
>{\columncolor[HTML]{FFFFFF}}c |
>{\columncolor[HTML]{FFFFFF}}c |}
\hline
\textbf{System}                      & \textbf{Disk} & \textbf{Ball (Solid)} & \textbf{Ball (Hollow)} \\ \hline
\textbf{Training time   {[}hr{]}}                   & 228.91            & 18.43                 & 15.31                  \\ \hline
\textbf{Mean Force Error}                      & 0.0004           & 0.0004
              & 0.0004              \\ \hline
\textbf{Energy MAE (\%)}                           & 0.0001            & 0.004                 & 0.005                  \\ \hline
\textbf{Lyap. Exponent}              & 0.4489          & 0.6082
               & 0.3583                \\ \hline
\textbf{Char. Time }                     & 2.2278           & 1.6441
              & 2.7906              \\ \hline
\end{tabular}}
\end{table}


\begin{table}[htb!]
\centering
\caption{Training times, energy mean absolute error (MAE), Lyapunov exponent and characteristic times for the experiments with noise.}
\label{GRWNoise}
\scalebox{0.85}{
\begin{tabular}{|
>{\columncolor[HTML]{FFFFFF}}c |
>{\columncolor[HTML]{FFFFFF}}c |
>{\columncolor[HTML]{FFFFFF}}c |
>{\columncolor[HTML]{FFFFFF}}c |
>{\columncolor[HTML]{FFFFFF}}c |}
\hline
\textbf{System}                      & \textbf{Disk} & \textbf{Ball (Solid)} & \textbf{Ball (Hollow)} \\ \hline
\textbf{Training time   {[}hr{]}}                  & 22.03         & 15.19                 & 15.06                  \\ \hline
\textbf{Mean Force Error}         
             & 0.0102           & 0.0132
              & 0.0196              \\ \hline
\textbf{Energy MAE (\%)}                          & 0.25          & 0.084                 & 0.134                  \\ \hline
\textbf{Lyap. Exponent }    & 0.5822     & 0.1407           & 0.4493                 \\ \hline
\textbf{Char. Time}                      & 1.7175        & 7.1063            & 2.2255
                 \\ \hline
\end{tabular}}
\end{table}

\begin{figure}[htbp]
    \centering
    \includegraphics[width=0.8\textwidth]{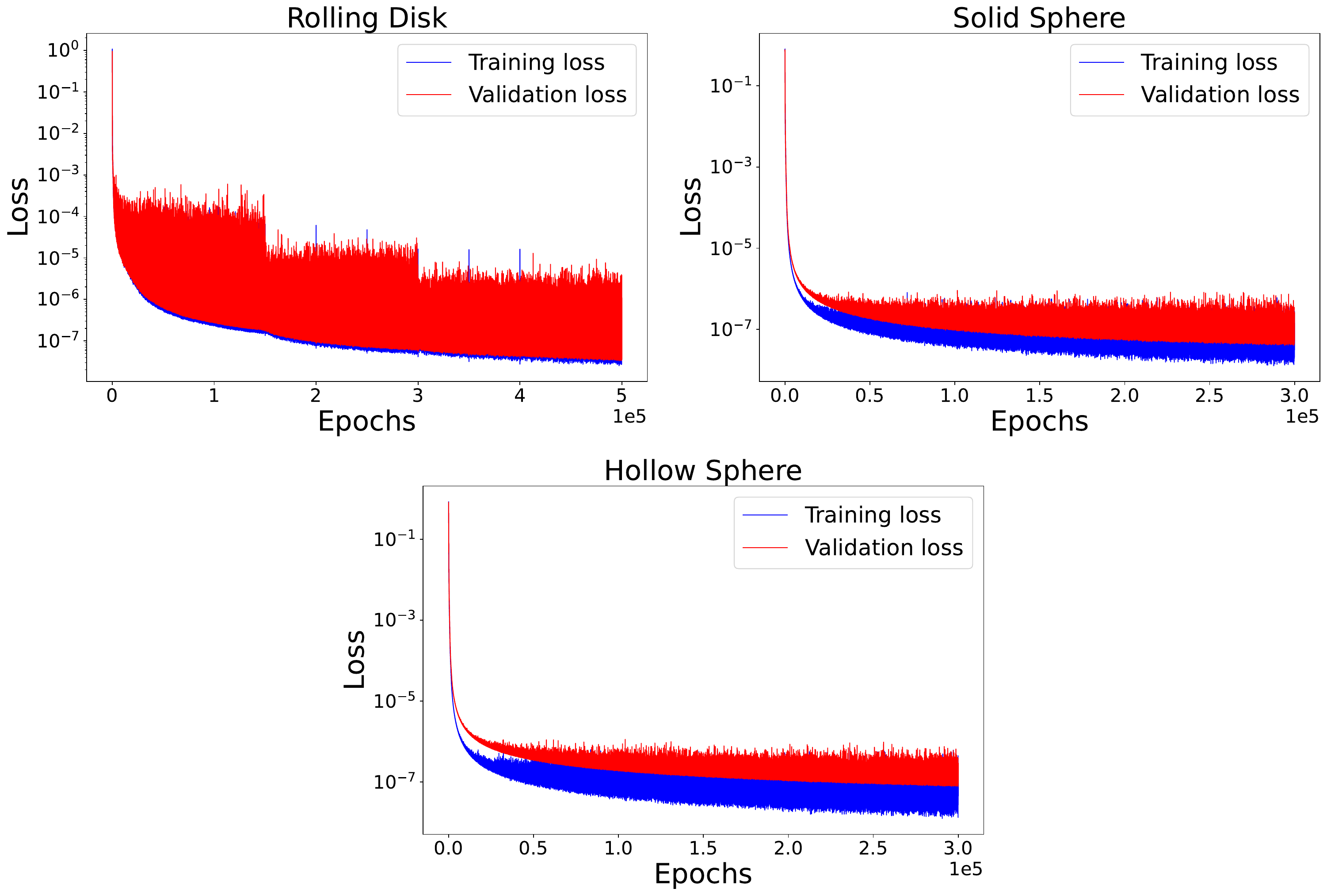}
    \caption{Evolution of the training and validation loss functions for the experiments without noise.} 
    \label{Losses}
\end{figure}
\begin{figure}[htbp]
    \centering
    \includegraphics[width=0.8\textwidth]{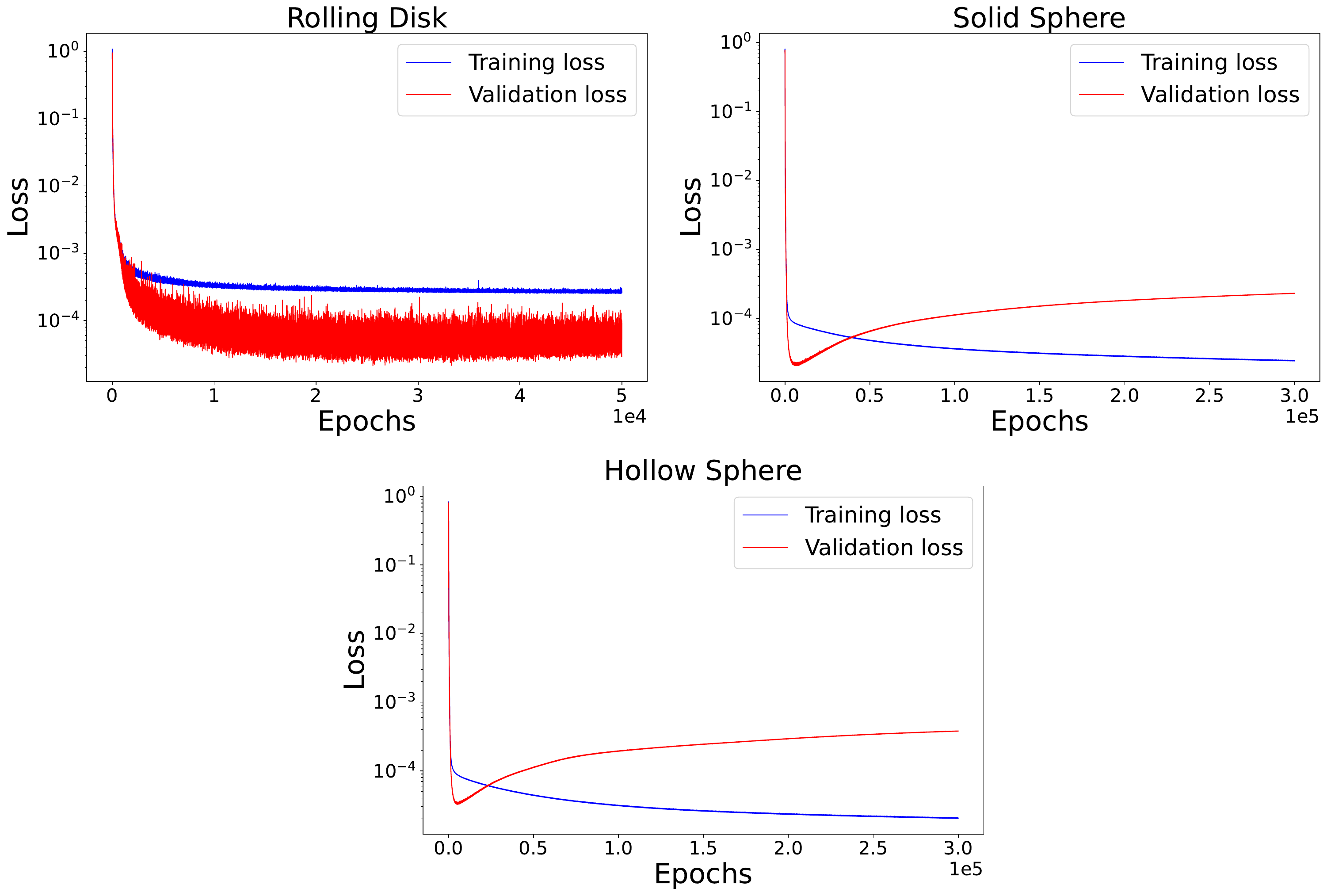}
    \caption{Evolution of the training and validation loss functions for the experiments with noise.} 
    \label{Lossesn}
\end{figure}

Figure \ref{Losses} shows the evolution of the training and validations loss functions for the experiments without noise. In particular, the effect of using different learning rates during the training process can be clearly seen for the rolling disk case. Figure \ref{Lossesn} shows the loss functions for the experiments with noise. The minima in the validation loss functions implies overfitting due to noise, being particularly prominent for both ball on table experiments. 

\clearpage
\section{Conclusion}

We propose a multi-network Hamiltonian-based architecture capable of modeling systems under holonomic and nonholonomic constraints from data. The novelty of our approach lies in the fact that that it simultaneously learns the behavior and structure of the system, providing a model for the Hamiltonian, constraints, and Lagrange multipliers without requiring explicit information on these quantities.\\

The rolling disk and ball on a spinning table are considered as examples of nonholonomic systems to asses the performance of the proposed approach. The architecture is shown to effectively recover the behavior of the system, predicting trajectories that closely follow the ground truth. Our approach is also shown to accurately and simultaneously model the Hamiltonian, constraints, and Lagrange multipliers of the system. When noise is introduced in the training set, the networks effectively recover the main behaviors of the system, and accurately model the constraints and their Lagrange multipliers.\\

Possible directions for future work include adaptations to obtain the form of the generalized momenta $\textbf{p}$ when only information on $\textbf{q}$ and $\dot{\textbf{q}}$ is available, extending the formulation to globally model nonlinear nonholonomic constraints, and assessing changes in performance when replacing neural networks by other modeling approaches.

\section*{Declaration of interest}
The authors have no conflicts to disclose.

\section*{Acknowledgments}
The authors wish to acknowledge the financial support of Natural Sciences and Engineering Research Councils
(NSERC) of Canada, under their Discovery Grant Program (No.
GR000989).
This research was supported in part through computational resources
and services provided by Advanced Research Computing at the University of British Columbia.

\section*{Declaration of generative AI and AI-assisted technologies in the writing process}

During the preparation of this work the authors used OpenAI's ChatGPT-4 to improve the readability and language of the abstract. After using this tool, the authors reviewed and edited the content as needed and take full responsibility for the content of the published article.

\bibliographystyle{elsarticle-num} 
\bibliography{cas-refs}





\end{document}


\begin{frontmatter}



\title{Hamiltonian-based neural networks for systems under nonholonomic constraints: Supplementary material}


\author[inst1]{Ignacio Puiggros T.}

\author[inst1]{A. Srikantha Phani}

\affiliation[inst1]{organization={Department of Mechanical Engineering, The University of British Columbia},
            addressline={2054-6250 Applied Science Lane}, 
            city={Vancouver},
            postcode={V6T 1Z4}, 
            state={British Columbia},
            country={Canada}}

\end{frontmatter}

\section{Nonholonomically constrained particle}

\subsection{System description}
As a proof of concept, we choose a simple system consisting of a particle moving in three dimensions under the effect of gravity and under a single nonholonomic constraint. The Hamiltonian of the system is 
\begin{equation} \label{H1}
    H=\frac{(p_{x}^{2}+p_{y}^{2}+p_{z}^{2})}{2m}+mgz
\end{equation}
where $g$ is the acceleration due to gravity and $m$ is the mass of the particle. The generalized momenta of the system correspond to the vector

\begin{equation} \label{p1}
    \textbf{p}=m\dot{\textbf{x}}
\end{equation}

where $\dot{\textbf{x}}$ is the velocity vector and $\textbf{p}$ is the generalized momenta. The system is under the nonholonomic constraint
\begin{equation} \label{c1}
    \dot{z}=y\dot{x}
\end{equation}


As the system has six dimensions in its phase space and there is a single constraint, the networks associated with the Hamiltonian and the multipliers take six inputs and give one output each, while the network assigned to learn the constraint matrix has three neurons in its input layer and three in its output layer, corresponding to the elements of $\textbf{A}$. Table \ref{TablePM} summarizes the parameters and architectures used for training for the point mass problem.\\

\begin{table}[htbp]
\centering
\caption{Parameters used for the point mass experiment.}
\label{TablePM}
\begin{tabular}{|l|ccc|}
\hline
\multicolumn{1}{|c|}{\textbf{Parameter}} & \multicolumn{1}{c|}{$H$} & \multicolumn{1}{c|}{$\bm{\lambda}$} & $\textbf{A}$ \\ \hline
Hidden layers                            & \multicolumn{1}{c|}{2}            & \multicolumn{1}{c|}{2}                  & 2            \\ \hline
Neurons per   layer                      & \multicolumn{1}{c|}{270}          & \multicolumn{1}{c|}{270}                & 50           \\ \hline
Activation   function                    & \multicolumn{1}{c|}{$Tanh$}         & \multicolumn{1}{c|}{$Tanh$}               & $Tanh$         \\ \hline
Learning rate                            & \multicolumn{1}{c|}{5e-5}         & \multicolumn{1}{c|}{5e-5}               & 3e-6         \\ \hline
Epochs                                   & \multicolumn{3}{c|}{3e5}                                                                   \\ \hline
Training set                             & \multicolumn{3}{c|}{5e3 points}                                                            \\ \hline
Validation set                           & \multicolumn{3}{c|}{2e3 points}                                                            \\ \hline
Batches                                  & \multicolumn{3}{c|}{1}                                                                    \\ \hline
\end{tabular}
\end{table}

The dataset was generated artificially by sampling points from independent uniform distributions in each coordinate in the range $[-3,3]$ and then imposing the constraint in \ref{c1}. To test the architecture, we use a sample trajectory with initial conditions $[1,1,1,1,1,1]$ in phase space.\\ 

As in the other experiments, the loss function used to simultaneously train the networks is:

\begin{equation} \label{NHHNNLoss}
   Loss=\sum_{i=1}^{N} \left( \dot{\textbf{q}}_{i}-\frac{\partial \hat{H}}{\partial \textbf{p}_{i}}\right)^{2}  + \left(\dot{\textbf{p}}_{i}+\frac{\partial \hat{H}}{\partial \textbf{q}_{i}} - \hat{\textbf{A}}^{T}\boldsymbol{\hat{\lambda}}\right)^{2} +
   \sum_{i=1}^{N} \left( \hat{\textbf{A}}\dot{\textbf{q}}_{i}-\boldsymbol{\hat{b}}\right)^{2}
\end{equation}

 \noindent where $\hat{H}$, $\hat{\textbf{A}}$,  $\boldsymbol{\hat{b}}$ and $\bm{\hat{\lambda}}$ are learned quantities and we sum over N points in the training set.\\

\subsection{Results without noise}

\begin{figure}[htbp]
    \centering
    \includegraphics[width=0.7\textwidth]{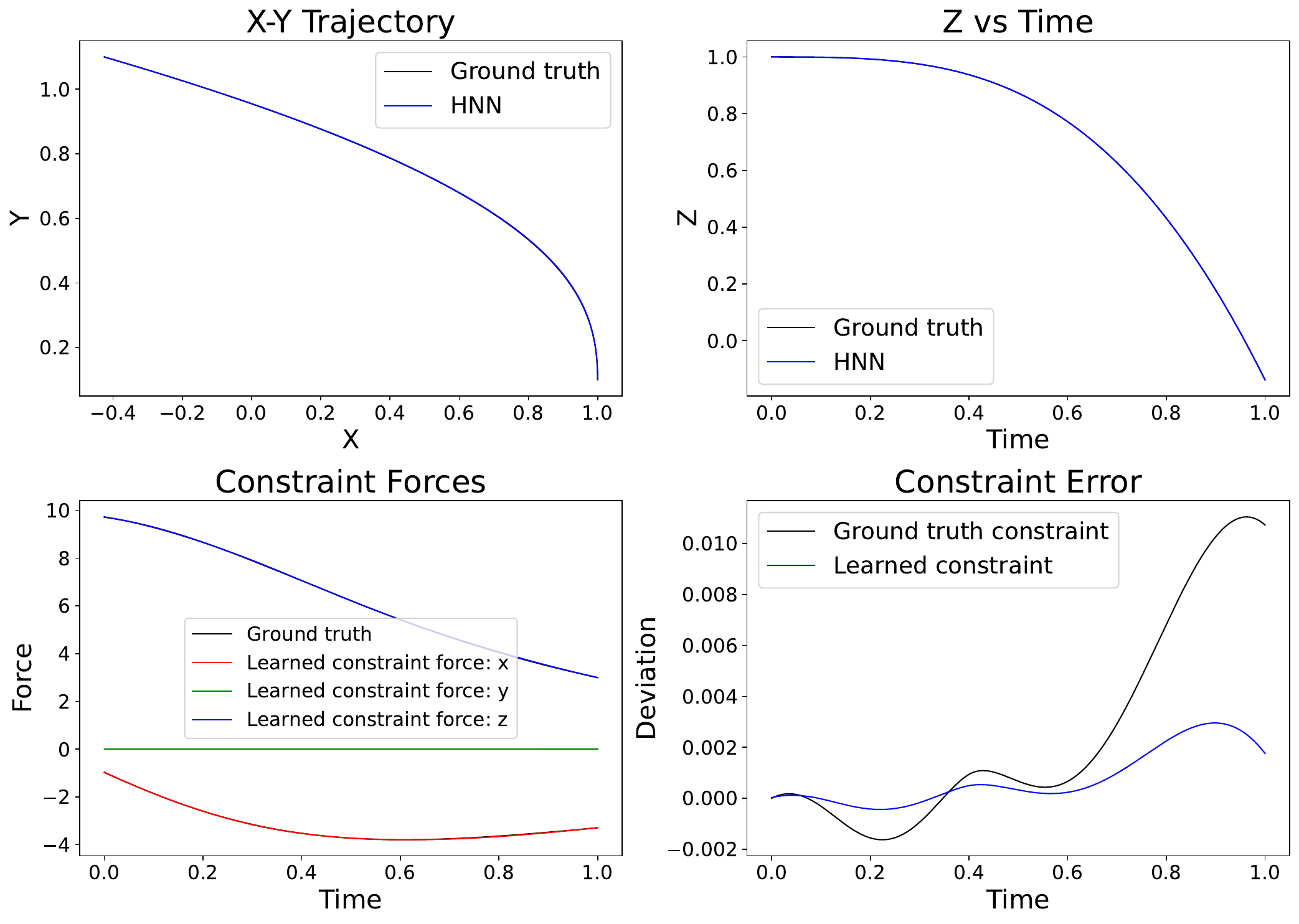}
    \caption{Trajectories and deviations for the nonholonomically constrained particle
.} 
    \label{PM_Traj}
\end{figure}

Top row of Figure \ref{PM_Traj} shows the top view for the test trajectory and the evolution of the $z$ coordinate in time. Bottom row shows the learned constraint forces on the left, corresponding to $\textbf{A}^{T}\boldsymbol{\lambda}$ and computed over the modeled trajectory. In order to asses how precisely the constraints hold, we can define the deviation from the analytical constraints as $\textbf{A} \dot{\boldsymbol{q}}-\boldsymbol{b}$, and $\hat{\textbf{A}}\dot{\textbf{q}}-\boldsymbol{\hat{b}}$ for the learned constraints, where $\dot{\textbf{q}}$ corresponds to the velocities of the modeled (not ground truth) trajectory. The deviation over time from both the learned and ground truth constraints is shown on the bottom right of Figure \ref{PM_Traj}. It can be seen that both the trajectory and the constraint forces match the ground truth with high precision. Both the learned and ground truth constraints deviations are close to zero, with the former being lower. This is probably due to the fact that the loss function in eq. \ref{NHHNNLoss} has a term that explicitly penalizes nonzero values for $\hat{\textbf{A}}\dot{\textbf{q}}-\boldsymbol{\hat{b}}$, while $\textbf{A}\dot{\textbf{q}}-\boldsymbol{b}$ is only minimized implicitly (i.e. it will be minimized as the correct trajectory is learned, but the term is not explicitly included in the loss function).\\

\subsection{Results with noise}

The experiment was repeated with a noisy dataset to assess the performance of the proposed architecture under imperfect information. A multiplicative noise model is used, where every variable $x$ in the data matrix is independently modified as:

\begin{equation}\label{noise}
    x \rightarrow x(1+\mathcal{N}(0,\sigma^{2}))
\end{equation}

\noindent where $\mathcal{N}(0,\sigma^{2})$ is a number sampled from a normal distribution with zero mean and a standard deviation of $\sigma=0.02$. We note that, in order to asses overfitting due to noise more effectively, we use eq. \ref{noise} on the training data but not to the validation set. In order to emulate the repeated measurements that are often taken in real experiments to deal with noise, ten iterations of the noisy dataset are computed (all from the same ``clean'' original dataset), and their average is used to train the networks.\\

The networks are trained exactly as in the case without noise. As with the previous experiment, the ground truth and learned constraint errors are defined as $\textbf{A} \dot{\boldsymbol{q}}-\boldsymbol{b}$ and $\hat{\textbf{A}}\dot{\textbf{q}}-\boldsymbol{\hat{b}}$ where $\boldsymbol{q}$ corresponds to the modeled trajectory.\\

\begin{figure}[htb!]
    \centering
    \includegraphics[width=0.7\textwidth]{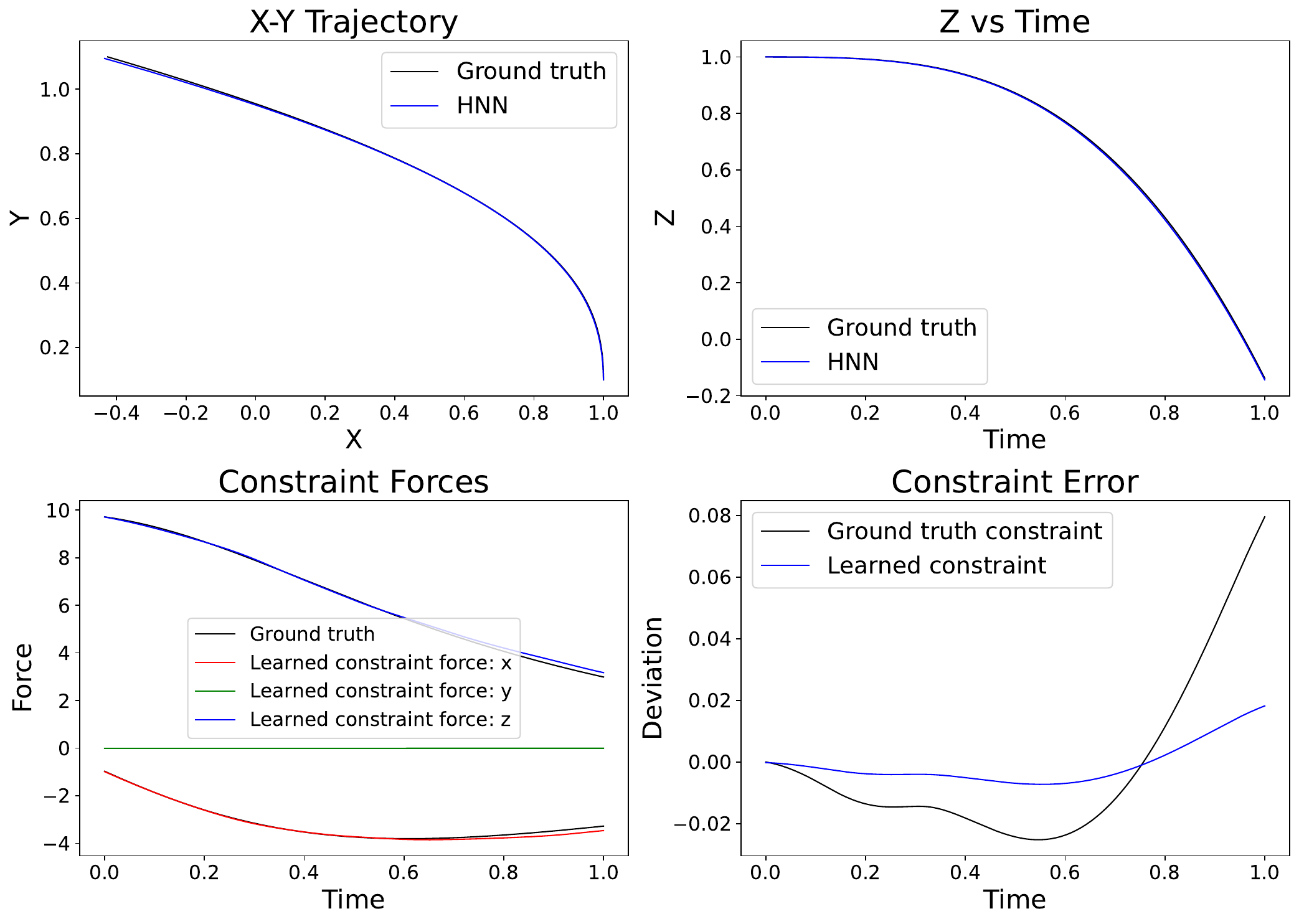}
    \caption{Trajectories and deviations for the point mass system in the presence of noise. } 
    \label{PM_Traj_noise}
\end{figure}

Top row of Figure \ref{PM_Traj_noise} shows the $x-y$ trajectory of the particle and the evolution of its $z$ coordinate over time. Bottom row shows the constraint forces along the trajectory and the deviations from the analytical and learned constraints. It can be seen that the trajectories, constraints and multipliers are recovered effectively, with deviations that are reasonable but higher than in the case without noise.

\subsection{General results}

In this subsection we summarize results corresponding to the training time, stability of the predicted trajectories over time, and precision of the learned Hamiltonian and constraint forces. The quantities shown are defined as described in the manuscript.\\



Table \ref{GRNoNoise} shows the results for the cases with and without noise. In the case without noise, both the energy and the forces are learned with high precision, which speaks of an effective recovery of the Hamiltonian, constraints, and multipliers of the system. Although the modeling error of the energy increases significantly in the presence of noise, this does not seem to be the case for the constraint forces, which maintain a low normalized deviation. This may suggest high robustness of the learned constraints and multipliers to imperfections in the training set. The Lyapunov exponents (and therefore the characteristic divergence times) are shown to be of the same order of magnitude across experiments, which speaks of the stability of the proposed architecture.


\begin{table}[htb!]
\centering
\caption{Training times, mean force error, mean absolute energy error, Lyapunov exponent and characteristic times for the nonhollonomically constrained particle.}
\label{GRNoNoise}
\scalebox{0.85}{
\begin{tabular}{|
>{\columncolor[HTML]{FFFFFF}}c |
>{\columncolor[HTML]{FFFFFF}}c |
>{\columncolor[HTML]{FFFFFF}}c |
>{\columncolor[HTML]{FFFFFF}}c |
>{\columncolor[HTML]{FFFFFF}}c |}
\hline
\textbf{System}                     & \textbf{No noise} & \textbf{Noise}\\ \hline
\textbf{Training time   {[}hr{]}}   & 7.29                & 7.97                         \\ \hline
\textbf{Mean Force Error}         & 0.0018             & 0.0125              \\ \hline
\textbf{Energy MAD (\%)}            & 0.076                & 0.303                          \\ \hline
\textbf{Lyap. Exponent   {[}1/s{]}} & 0.2047             & 0.2118            \\ \hline
\textbf{Char. Time {[}s{]}}         & 4.8849             & 4.7205          \\ \hline
\end{tabular}}
\end{table}

\begin{figure}[htbp]
    \centering
    \includegraphics[width=1\textwidth]{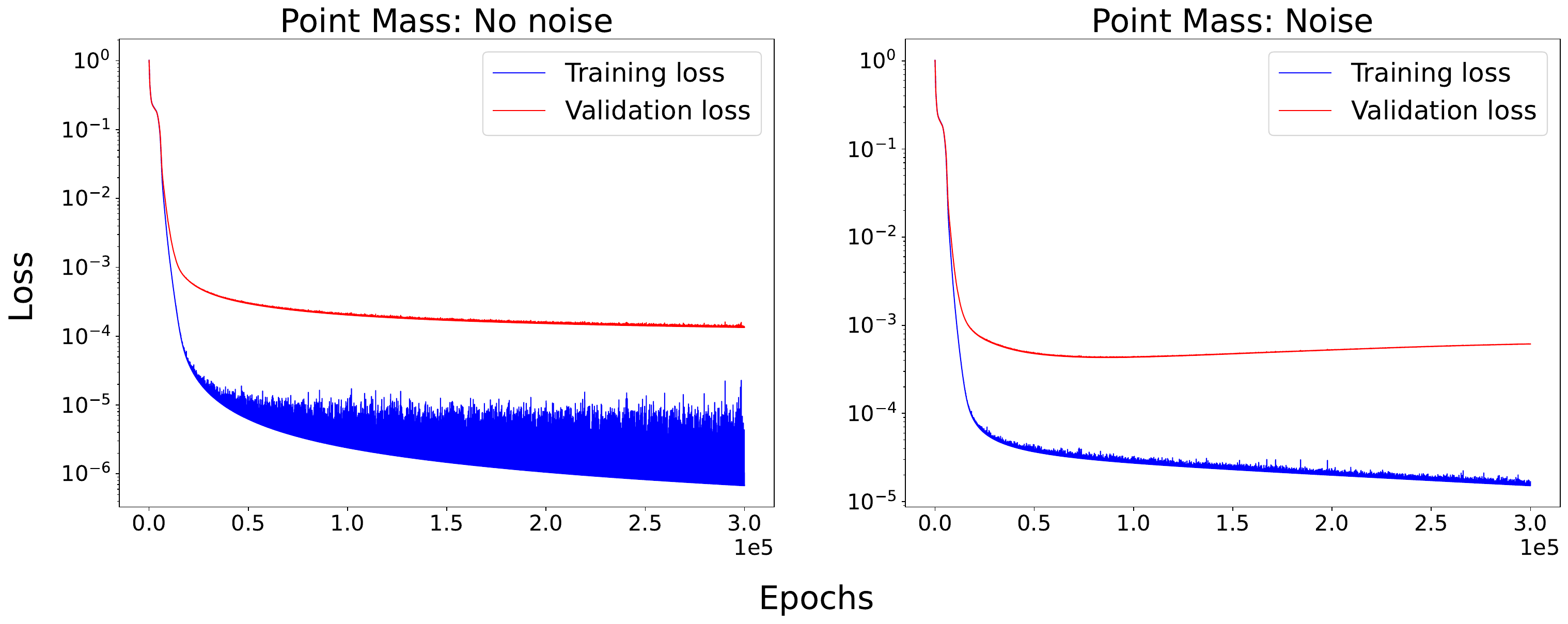}
    \caption{Evolution of the training and validation loss functions for the nonholonomically constrained particle.} 
    \label{LossesSI}
\end{figure}

Left side of Figure \ref{LossesSI} shows the evolution of the training and validations loss functions for the experiment without noise, while the right side shows the case with noise. The validation loss function shows a minima around $7.5e4$ epochs, which implies overfitting due to noise after this point.

\clearpage

\section{Uniqueness of $\hat{H}$ and $\boldsymbol{\hat{\lambda}}$}
\label{sec:sample:appendix}
A relevant question is whether the three networks will learn exactly and only the quantity they are assigned to. In principle, the networks could learn in such a way that both $\dot{\textbf{q}}$ and $\dot{\textbf{p}}$ are properly modeled, but the terms $\hat{H}$ and $\hat{\textbf{A}}^{T}\boldsymbol{\hat{\lambda}}$ do not correspond to the actual Hamiltonian and constraint forces. If this were to occur, we would have that the networks ``mixed'' the learned quantities, i.e. $\hat{H}$ may include extra terms corresponding to the constrained dynamics, and/or $\hat{\textbf{A}}^{T}\boldsymbol{\hat{\lambda}}$ may incorporate part of the symplectic dynamics, both of which are undesirable.\\

Looking at the loss function in eq. \ref{NHHNNLoss}, we see that to minimize the first half of the first summation, the terms of $\hat{H}$ that depend on $\textbf{p}$ should converge to the corresponding terms in the actual Hamiltonian $H$. Similarly, as the loss function is minimized, the second sum guarantees that $\hat{\textbf{A}}$ and $\boldsymbol{\hat{b}}$ will converge to a possible representation of the constraints of the system as long as they are nonzero. On the other hand, the second term of the first sum only guarantees that 
$\frac{\partial \hat{H}}{\partial \textbf{q}} - \hat{\textbf{A}}^{T}\boldsymbol{\hat{\lambda}}$ will converge to $\dot{\textbf{p}}$, but nothing is explicitly imposed on $\boldsymbol{\hat{\lambda}}$ or on the terms of $\hat{H}$ that depend on $\textbf{q}$. Therefore, the question is if $\hat{H}$ and $\boldsymbol{\hat{\lambda}}$ will converge to the desired quantities.\\

To conclude that $\hat{H}$ and $\boldsymbol{\hat{\lambda}}$ will converge to the desired quantities, it is sufficient to show that $H$ and $\boldsymbol{\lambda}$ have a unique solution (up to a constant) for the equation 

\begin{equation}\label{pdot}
\dot{\textbf{p}}=\frac{\partial H}{\partial \textbf{q}} - \textbf{A}^{T}\boldsymbol{\lambda}
\end{equation}

\noindent for given $\textbf{A}$ and $\dot{\textbf{p}}$.\\

It will be shown for the three dimensional case that there is a sufficient condition on the matrix $\textbf{A}$ that guarantees the uniqueness of the solution of $H$ and $\boldsymbol{\lambda}$, for given $\textbf{A}$ and $\dot{\textbf{p}}$. \\

Let $H_{1}$ and $\boldsymbol{\lambda}_{1}$ be a solution of eq. \ref{pdot} given $\textbf{A}$ and $\dot{\textbf{p}}$. If a different set of solutions $H_{2}$ and $\boldsymbol{\lambda}_{2}$ exists, then we may define $H^{*}\equiv H_{1}-H_{2}$ and $\boldsymbol{\lambda}^{*}\equiv \boldsymbol{\lambda}_{1}-\boldsymbol{\lambda}_{2}$. Using the fact that $H_{2}$ and $\boldsymbol{\lambda}_{2}$ are solutions of eq. \ref{pdot} we have

\begin{equation}\label{pdot2}
\begin{split}
\dot{\textbf{p}}&=\frac{\partial H_{2}}{\partial \textbf{q}} - \textbf{A}^{T}\boldsymbol{\lambda}_{2}\\
&=\frac{\partial (H_{1} - H^{*})}{\partial \textbf{q}} - \textbf{A}^{T}(\boldsymbol{\lambda}_{1}-\boldsymbol{\lambda}^{*})\\
&=\frac{\partial H_{1} }{\partial \textbf{q}} - \textbf{A}^{T}\boldsymbol{\lambda}_{1}-\left(\frac{\partial H^{*}}{\partial \textbf{q}}-\textbf{A}^{T}\boldsymbol{\lambda}^{*}\right)\\
\end{split}
\end{equation}

 In the last line, it can immediately be seen that the left hand side cancels with the first two terms in the right hand side as $H_{1}$ and $\boldsymbol{\lambda}_{1}$ are solutions of eq. \ref{pdot}. Therefore, we get that for $H_{2}$ and $\boldsymbol{\lambda}_{2}$ to be different solutions to eq. \ref{pdot}, we must have

\begin{equation}\label{condition1}
\frac{\partial H^{*}}{\partial \textbf{q}}=\textbf{A}^{T}\boldsymbol{\lambda}^{*}
\end{equation}

 Which is only possible if the term $\textbf{A}^{T}\boldsymbol{\lambda}^{*}$ is a curless field, as the left hand side is curless due to being a gradient. We then have the following as a necessary condition for $H_{2}$ and $\boldsymbol{\lambda}_{2}$ to be different solutions to eq. \ref{pdot}

 \begin{equation}\label{condition2}
    \nabla \times (\textbf{A}^{T}\boldsymbol{\lambda}^{*})=0
 \end{equation}

Therefore, uniqueness is guaranteed as long as eq. \ref{condition2} lacks a non trivial solution for $\boldsymbol{\lambda}^{*}$. \\

Note that if uniqueness is shown for some representation of the constraints $\textbf{A}$, it will also be true for any other representation obtained from the transformations described in the main text, as $\boldsymbol{\lambda}^{*}$ transforms in such a way that $\textbf{A}^{T}\boldsymbol{\lambda}^{*}$ remains invariant.\\

As an example, lets consider the simple point mass case, with the Hamiltonian specified in eq. \ref{H1}, generalized momenta as in eq. \ref{p1}, and under the nonholonomic constraint in eq. \ref{c1}. This constraint corresponds to the matrix

\begin{equation} \label{A}
    \textbf{A}=[-y, 0, 1]
\end{equation}

We then have that eq. \ref{condition2} becomes

\begin{equation} \label{A}
\nabla \times (\textbf{A}^{T}\lambda^{*})=
\nabla \times
\begin{pmatrix} 
-y\lambda^{*} \\ 0 \\ \lambda^{*}
\end{pmatrix}
=
\begin{pmatrix} 
\frac{\partial \lambda^{*}}{\partial y} \\
-y\frac{\partial \lambda^{*}}{\partial z}-\frac{\partial \lambda^{*}}{\partial x}
\\ 
y\frac{\partial \lambda^{*}}{\partial y}+\lambda
\end{pmatrix}
=0
\end{equation}
which is only solved by $\lambda^{*}=0$. This implies that the solutions for $H$ and $\lambda$ are unique, and therefore, that the networks will learn the desired quantities as they converge.